\documentclass[10pt]{elsarticle}

\usepackage[utf8]{inputenc}
\usepackage{graphicx} 
\usepackage{caption}
\usepackage{enumitem}
\usepackage{algorithm}
\usepackage[noend]{algpseudocode}
\usepackage{subcaption}
\usepackage{appendix}
\usepackage{anysize}
\usepackage{mathtools}
\usepackage{soul}
\usepackage{placeins}
\usepackage[rightcaption]{sidecap}
\usepackage{nomencl}
\usepackage{wrapfig}
\usepackage{multicol}
\usepackage[letterpaper,top=2cm,bottom=2cm,left=3cm,right=3cm,marginparwidth=1.75cm]{geometry}
\usepackage{comment}

\usepackage{amsmath}
\usepackage{amssymb}
\usepackage{amsthm}
\usepackage{cancel}
\usepackage{graphicx}
\usepackage{booktabs}
\usepackage{xspace} 
\usepackage{nicematrix}
\usepackage{scalerel,stackengine}
\usepackage[colorlinks=true, allcolors=blue]{hyperref}
\usepackage{tikz}

\usetikzlibrary{fit}
\usetikzlibrary{positioning}
\usetikzlibrary{calc}

\newcommand{\coo}{\ensuremath{\mathrm{CO_2}}\xspace}
\newcommand{\rd}{\mathrm{d}}
\newcommand{\pd}[2]{\frac{\partial #1}{\partial #2}}
\newcommand{\dd}[2]{\frac{\rd #1}{\rd #2}}

\newcommand{\cU}{\mathcal{U}}
\newcommand{\EulerFlux}{\mathcal{F}}
\newcommand{\Source}{\mathcal{S}}

\newcommand{\Vol}{v}
\newcommand{\hllcF}{F}
\newcommand{\f}{\psi}
\newcommand{\fg}{\psi_g}
\newcommand{\fl}{\psi_l}
\newcommand{\tRho}{\ensuremath{\Tilde{\rho}}}
\newcommand{\te}{\ensuremath{\Tilde{e}}}
\newcommand{\fOne}{f_1}
\newcommand{\fTwo}{f_2}
\newcommand{\fThree}{f_3}

\newcommand{\Hammer}{Hammer et al.\space}
\newcommand{\dt}[1]{\(\Delta t = #1\)~s}

\newcommand{\fullEqns}{\text{full VLE equations}\xspace}
\newcommand{\reducedDae}{\textit{Reduced-VLE-Algebraic}\xspace}
\newcommand{\fullDae}{\textit{Full-VLE-Algebraic}\xspace}
\newcommand{\ode}{\textit{Reduced-VLE-ODE}\xspace}
\newcommand{\EvaluatedAt}[2]{\left( #1 \right)_{#2}}
\newcommand{\odt}{\(\mathcal{O}({\Delta t})\)}
\newcommand{\odtt}{\(\mathcal{O}({\Delta t^2})\)}

\theoremstyle{plain}
\newtheorem{remark}{Remark}

\allowdisplaybreaks
\stackMath

\graphicspath{{images/}}

\title{A new temperature evolution equation that enforces thermodynamic vapour-liquid equilibrium in multiphase flows - application to \coo modeling.}

\makenomenclature

\begin{document}


\begin{frontmatter}


\author[cwi,delft]{Pardeep Kumar\corref{cor1}}
\ead{pardeep@cwi.nl}

\author[cwi,eindhoven]{Benjamin Sanderse}

\author[shell]{Patricio I. Rosen Esquivel}

\author[shell,delft]{R.A.W.M. Henkes}

\cortext[cor1]{Corresponding author}

\affiliation[cwi]{
    organization={Centrum Wiskunde \& Informatica},
    city={Amsterdam},
    country={The Netherlands}
}
\affiliation[eindhoven]{
    organization={Eindhoven University of Technology},
    city={Eindhoven},
    country={The Netherlands}
}
\affiliation[shell]{
    organization={Shell Projects and Technology},
    city={Amsterdam},
    country={The Netherlands}
}
\affiliation[delft]{
    organization={Delft University of Technology},
    city={Delft},
    country={The Netherlands}
}

\begin{abstract}

This work presents a novel framework for numerically simulating the depressurization of tanks and pipelines containing carbon dioxide (\coo). The framework focuses on efficient solution strategies for the coupled system of fluid flow equations and thermodynamic constraints. A key contribution lies in proposing a new set of equations for phase equilibrium calculations which simplifies the traditional vapor-liquid equilibrium (VLE) calculations for two-phase \coo mixtures. The first major novelty resides in the reduction of the conventional four-equation VLE system to a single equation, enabling efficient solution using a non-linear solver. This significantly reduces computational cost compared to traditional methods. Furthermore, a second novelty is introduced by deriving an ordinary differential equation (ODE) directly from the UV-Flash equation. This ODE can be integrated alongside the governing fluid flow equations, offering a computationally efficient approach for simulating depressurization processes.

\end{abstract}




\end{frontmatter}

\section{Introduction}

Carbon capture and storage (CCS) is a promising feasible alternative for mitigating greenhouse gas emissions. In CCS, \coo needs to be transported and conditioned from its capture location to a storage facility. This process is normally accomplished via pipelines and ships. For short distances and small volumes, transporting \coo in gaseous or liquid form via ships can be cost-effective, but pipeline transport in a dense liquid-like state is a more economical and scalable option for large volumes and long distances \cite{noauthor_iea_2023}. The \coo is then re-injected through a well into the target storage location such as an aquifer or a depleted gas reservoir. The transport of \coo along pipelines and wells, including multiphase flow transport and its associated transients, is the focus of the current study.

During regular pipeline operation, changes in pressure and temperature along the pipeline can lead to multiphase flow behavior, such as gas-liquid (two-phase flow) or gas-liquid-solid (three-phase flow). Similarly, during pipeline depressurization, \coo will exhibit two-phase behaviour with rapid cooling along the saturation line. Predicting the lowest and highest temperatures during such an operation is crucial to design, control, and optimize the pipeline and manage its integrity \cite{hammer_method_2013}. For this purpose and many other applications, simulation tools offer an attractive alternative for predicting the performance of a given system under varying conditions. 

Two-phase flow of \coo is governed by the Navier-Stokes equations. A comprehensive overview of various two-phase flow models is available in \cite{bruce_stewart_two-phase_1984}. For an in-depth exploration of the models utilized in simulating \coo transport and experimental data, we recommend the review paper by Munkejord et al. \cite{munkejord_co2_2016} and the references therein. In pipeline applications, cross-sectional averaging is employed to simplify the Navier-Stokes equations and derive a one-dimensional two-fluid model \cite{ishii_thermo-fluid_2011}. However, this averaging procedure can render the model ill-posed, leading to discontinuous dependency of the solution on initial data and unbounded growth rates for the smallest wavelengths \cite{bruce_stewart_two-phase_1984, toumi_approximate_1999}. In this paper, we circumvent this issue by considering the unconditionally hyperbolic Homogeneous Equilibrium Model (HEM) \cite{lund_depressurization_2011}, which is suited for well-mixed two-phase flow systems in which the different phases travel at approximately the same velocity. It has been used in \coo simulation studies in for example \cite{giljarhus_solution_2012, hammer_method_2013, munkejord_depressurization_2015}. Additionally, it has found application in simulating heat exchangers and nuclear reactors \cite{toumi_weak_1991}. A comprehensive treatment of HEM is provided in the review articles by Stewart and Wendroff \cite{bruce_stewart_two-phase_1984} and Menikoff and Plohr \cite{menikoff_riemann_1989}.

While HEM offers simplicity, it is acknowledged for its limitations in handling strong non-equilibrium effects, such as those arising from droplets in gas flow. These effects, albeit correctable with additional terms, pose challenges to the model's accuracy. To effectively capture significant non-equilibrium effects, particularly of kinetic nature, more complex two-fluid models are warranted, accounting for the momentum of each phase \cite{clerc_numerical_2000}. Despite their computational expense, these models offer enhanced accuracy. However, our present investigation confines itself to HEM due to its simplicity.

Even though the HEM is a hyperbolic model, the fact that \coo is a non-ideal gas  complicates the simulation of two-phase \coo transport problems: in addition to the flow equations, one needs to solve an algebraic system of equations that describes the thermodynamic vapour-liquid equilibrium (VLE). As the evolution of density and internal energy is available through the fluid flow conservation equations, it is natural to solve the equilibrium problem in UV space \cite{smejkal_phase_2017} (where U denotes the internal energy, and V the inverse of the density). A few papers considering this so-called ``UV flash" are available in literature, see e.g.\ \cite{saha_isoenergetic-isochoric_1997, castier_solution_2009, michelsen_isothermal_nodate}, and the review of Smejkal et al. \cite{smejkal_phase_2017}. These UV equilibrium calculations consist of solving a system of non-linear algebraic equations. For a pure component fluid, this amounts to solving a system of four algebraic equations \cite{michelsen_thermodynamic_2007}, in each grid cell along the pipeline. The resulting coupled system of fluid flow and thermodynamics can be rather expensive to solve, especially when complex equations of state like Span-Wagner \cite{span_new_1996} are involved (which we will use in this work). 

In order to reduce the computational expense of solving a coupled system of dynamic flow equations with non-linear algebraic equations, we propose two new approaches. In the first approach, we reduce the four algebraic equations to a single equation by using saturation relations. This reduces the computational expense significantly, even though it still requires a non-linear equation solver. In the second approach, we derive an ordinary differential equation from the reduced representation and thereby eliminate the need of a non-linear solver, further reducing the computational expenses. To assess the new approaches, we utilize the forward Euler method for time integration \cite{hairer_numerical_1989}. 

We evaluate these algorithms by simulating the depressurization of a pipe filled with \coo. To gain fundamental insights into the role of the VLE on the algorithms, we initially test them on the depressurization of a tank filled with \coo. This is a simplified problem that shares several physical characteristics with pipeline depressurization. In this context, a relevant work is that of Sirianni et al.\ \cite{sirianni_explicit_2024}, who introduced an explicit conservative-primitive solver for general thermodynamic variables. However, their derivation relies on the choice of a specific time-integration method (forward Euler in their case). In contrast, our approach differs by focusing on discretizing the equations in space, leading to the derivation of an evolution equation for the temperature that is still continuous in time. In addition, the work by Sirianni et al.\ is limited to single-phase scenarios, whereas our methodology covers the case of two-phase systems as well.

The organization of this paper is as follows. In section \ref{sec:governing_equations}, we start by recapitulating the fluid model (HEM) and the constitutive relations from thermodynamics. In section \ref{sec:redcued_eqns}, we derive a new algebraic equation, followed by a new evolution equation for the temperature. In section \ref{sec:numerical_method}, we present spatial and temporal discretization schemes. In section \ref{sec:results}, we present results for two problems, namely the tank depressurization and the pipeline depressurization. Finally conclusions are summarized in section \ref{sec:conclusions}.  

\section{Governing equations}\label{sec:governing_equations}

\subsection{Fluid flow in a pipeline}\label{sec:HEM}
The Navier-Stokes (NS) equations govern the dynamics of fluid flow. For fully-dispersed two-phase fluid flow, the NS equations can be averaged and written in terms of mixture quantities, resulting in the Homogeneous Equilibrium Model (HEM). For flow through a horizontal pipe, the HEM  (without wall friction) takes the following form \cite{lund_depressurization_2011}:
\begin{equation}
\label{vectorised-model}
\partial_{t} \cU + \partial_{x} \EulerFlux(\cU) = \Source(\cU),
\end{equation}
where $\cU(x,t)$ is the vector of conserved variables, $\EulerFlux(\cU)$ the flux vector and $\Source(\cU)$ the source term, which have the following form:
\begin{equation}
\label{vector_eqns}
\cU = \begin{bmatrix}
{\rho }\\
{\rho u }\\
{\rho E}
\end{bmatrix}, \quad
\EulerFlux(\cU) = \begin{bmatrix}
{\rho u}\\
{\rho u^{2} + p }\\
{(\rho E + p) u}
\end{bmatrix} = \begin{bmatrix}
    \cU_{2} \\ \frac{\cU_{2}^2}{\cU_{1}} + p(\cU) \\ (\cU_{3} + p(\cU)) \frac{\cU_{2}}{\cU_{1}}
\end{bmatrix},
\quad
\Source(\cU) = \begin{bmatrix}
{0}\\
{0}\\
{0}
\end{bmatrix}.
\end{equation}
The model is supplemented with the constraint $\alpha_{g} + \alpha_{l} = 1$, where $\alpha_{g}(x,t)$ and $\alpha_{l}(x,t)$ are the gas and liquid volume fraction, respectively. The mixture density and energy are defined as $\rho := \alpha_{g} \rho_{g} + \alpha_{l} \rho_{l}$ and $\rho E :=\alpha_g \rho_g E_{g} + \alpha_l \rho_l E_{l}$, where
\begin{equation}
E_{k} = e_{k} + \frac{1}{2} u^2.
\end{equation}
Here, $\rho_{k}, e_k, E_k$ are the density, specific internal energy, and specific energy (kinetic + internal) of each phase. The mixture specific internal energy $e$ follows as $e = (\alpha_{g} \rho_{g} e_{g} + \alpha_{l} \rho_{l} e_{l})/\rho$. The two phases are assumed to have the same velocity $u$ and equilibrium thermodynamic pressure $p$. 

To close the system of fluid flow equations \eqref{vectorised-model}, the pressure $p(\cU)$ must be specified using an equation of state (EOS), as will be detailed in section \ref{sec:thermodynamics}. In addition, initial and boundary conditions are needed to complete the problem specification, these will be described in later sections.

The HEM model, when discretized with a finite volume method, leads to the following semi-discrete form for the $i^{th}$ finite volume:
\begin{align} \label{eqn:semiDiscrete}
\frac{\rd U_i}{\rd t} = -\frac{1}{\triangle x} (\Hat{\EulerFlux}_{i+\frac{1}{2}} - \Hat{\EulerFlux}_{i-\frac{1}{2}}), \qquad i=1\ldots N,
\end{align}
where $U_i(t) \in \mathbb{R}^{3} \approx \cU(x_{i},t)$, ${\triangle x}$ is the grid spacing, $\Hat{\EulerFlux}_{i\pm\frac{1}{2}}$ represent numerical fluxes and $N$ is the number of finite volumes. We use the Harten-Lax-van Leer-Contact(HLLC) scheme (see appendix \ref{HEM_HLLC}) to compute the numerical fluxes, $\Hat{\EulerFlux}_{i\pm\frac{1}{2}}$.  Equation \eqref{eqn:semiDiscrete} can be written in the general form 
\begin{equation}\label{eqn:semiDiscrete2}
    \frac{\rd U(t)}{\rd t} = f(U(t)),
\end{equation}
where
\begin{align*}
  U(t) = [U_{1}(t), \ldots, U_{i}(t), \ldots, U_{N}(t)]^T, \qquad U_{i} = [\rho_i, (\rho u)_i, (\rho E)_i]^T,
\end{align*}
and we have assumed that an expression of the form $p(U)$ is given by thermodynamics relations. This will be revisited in section \ref{sec:thermodynamics}. 

\subsection{Tank model}\label{sec:tank_equations}
As a simplification of the pipeline depressurization problem, we first consider the depressurization of a tank, previously studied for example in \cite{giljarhus_solution_2012,hammer_method_2014}. We can roughly think of it as a pipe discretized with a single cell. The governing equations for mass and internal energy can be written as:
\begin{equation}
\label{tank_eqns}
\begin{aligned}
    \frac{\rd \rho}{\rd t} &= - \frac{\dot{m}(\rho, T)}{\Vol},\\
    \frac{\rd (\rho e)}{\rd t} &= \frac{\dot{Q}(T) - \dot{m}(\rho, T) h(\rho, T)}{\Vol},
\end{aligned}
\end{equation}
where $\dot{m}$ is the mass flow rate and $\dot{Q}$ is the heat transfer rate, which for the tank problem are given by
\begin{align*}
    \dot{Q}(T) &= \eta A (T_{\mathrm{amb}} - T),\\
    \dot{m}(\rho, T) &= K_{v}\sqrt{\rho(p(\rho, T) - p_{\mathrm{amb}})}.
\end{align*}
Here, $p$, $T$, $\rho$, $e$ and $h$ denote the pressure, temperature, density, internal energy and enthalpy of \coo in the tank; $\Vol$ is the volume of the tank; $\eta$ the heat transfer coefficient between tank and ambient; $A$ is the surface area of the tank; $K_{v}$ the flow coefficient of the valve. The subscript `amb' denotes the ambient conditions outside the tank. 
Similar to the fluid flow equations \eqref{vectorised-model}, the governing equations can be expressed in the form:
\begin{align}
\frac{\mathrm{d} U}{\mathrm{d} t}=f(U),
\end{align}
where $U=[\rho,\rho e]^T$ and
\begin{align} \label{tank_eqn_dae_rhs}
    f(U) =  \frac{1}{\Vol}\begin{bmatrix} - \dot{m}(\rho, T) \\ \dot{Q}(T) - \dot{m}(\rho, T) h(\rho, T) \end{bmatrix}.
\end{align}

The system requires closure with thermodynamic relations that establish the connection between $p$, $T$ and $h$ and the conserved variables $\rho$ and $\rho e$, which will now be described.

\subsection{Thermodynamics and standard UV-Flash overview}\label{sec:thermodynamics}
For simple (e.g.\ ideal) gases, closed-form expressions for $p$ are sometimes available, such as $p=p(\rho,e)$ or $p=p(\rho,T)$ relations. For more complicated fluids like \coo, we need to use a real gas equation of state (EOS). The Span-Wagner EOS \cite{span_new_1996} is generally considered to be a very accurate description for pure \coo and will be employed here. We consider the formulation of the EOS specified in terms of the Helmholtz free energy $\mathcal{A}(\rho, T)$, where $\mathcal{A} = e - T s$ and $s$ is the specific entropy. All thermodynamic properties can be computed through derivatives of $\mathcal{A}(\rho, T)$, for example: 
\begin{align}
    p(\rho, T) &= \rho^2 \left(\frac{\partial \mathcal{A}}{\partial \rho}\right)_T, \label{p-hemholtz} \\
    e(\rho, T) &= \mathcal{A} - T\left(\frac{\partial \mathcal{A}}{\partial T}\right)_\rho, \label{e-hemholtz}\\
    s(\rho, T) &= -\left(\frac{\partial \mathcal{A}}{\partial T}\right)_\rho \label{s-hemholtz},\\
    h(\rho, T) &= \mathcal{A} + \rho \left(\frac{\partial \mathcal{A}}{\partial \rho}\right)_T -T\left(\frac{\partial \mathcal{A}}{\partial T}\right)_\rho \label{h-hemholtz}.
\end{align}
Given such an EOS, and two known thermodynamic quantities, one can solve for any of the other quantities. 

\begin{figure}[htbp]
     \centering
     \begin{subfigure}[b]{0.48\textwidth}
         \centering
         \includegraphics[width=\textwidth]{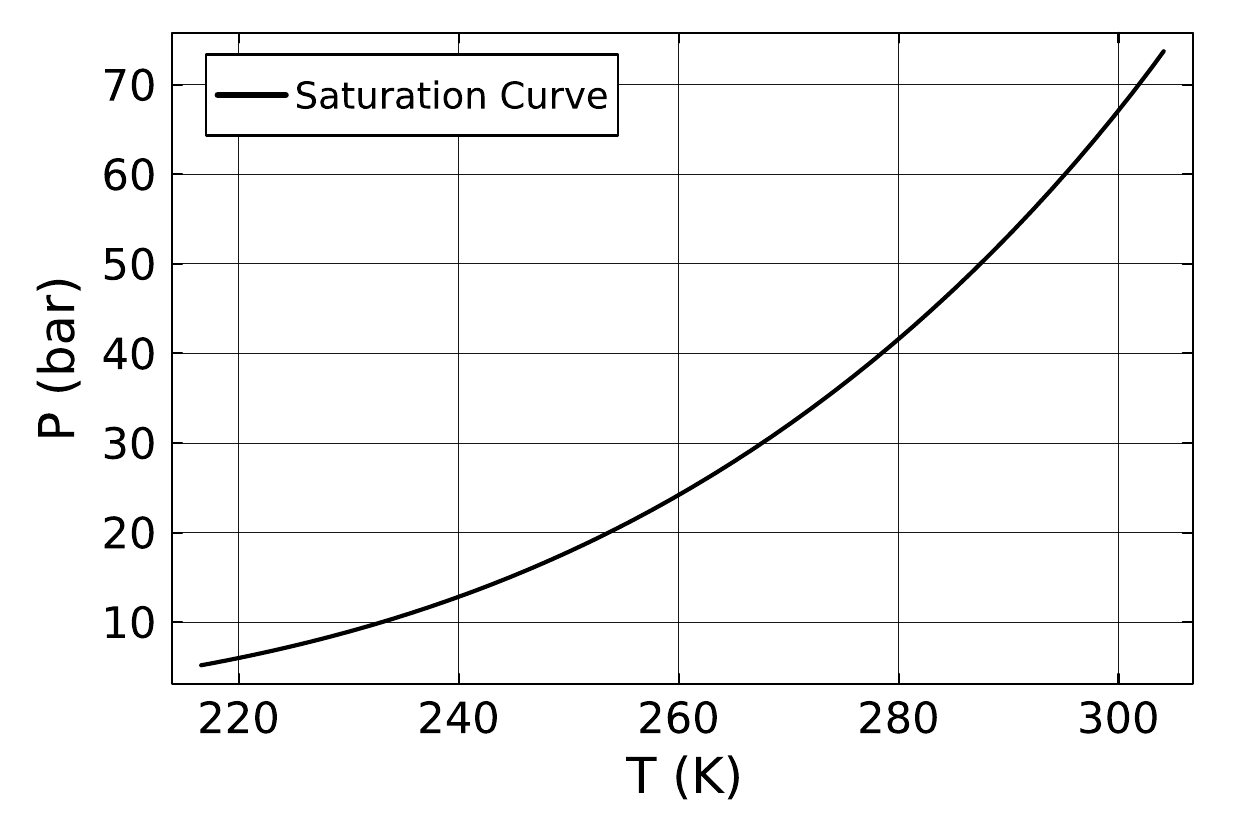}
         \caption{Saturation curve in $p-T$ co-ordinates}
         \label{fig:SaturationCurve_PT}
     \end{subfigure}
     \begin{subfigure}[b]{0.48\textwidth}
         \centering
         \includegraphics[width=\textwidth]{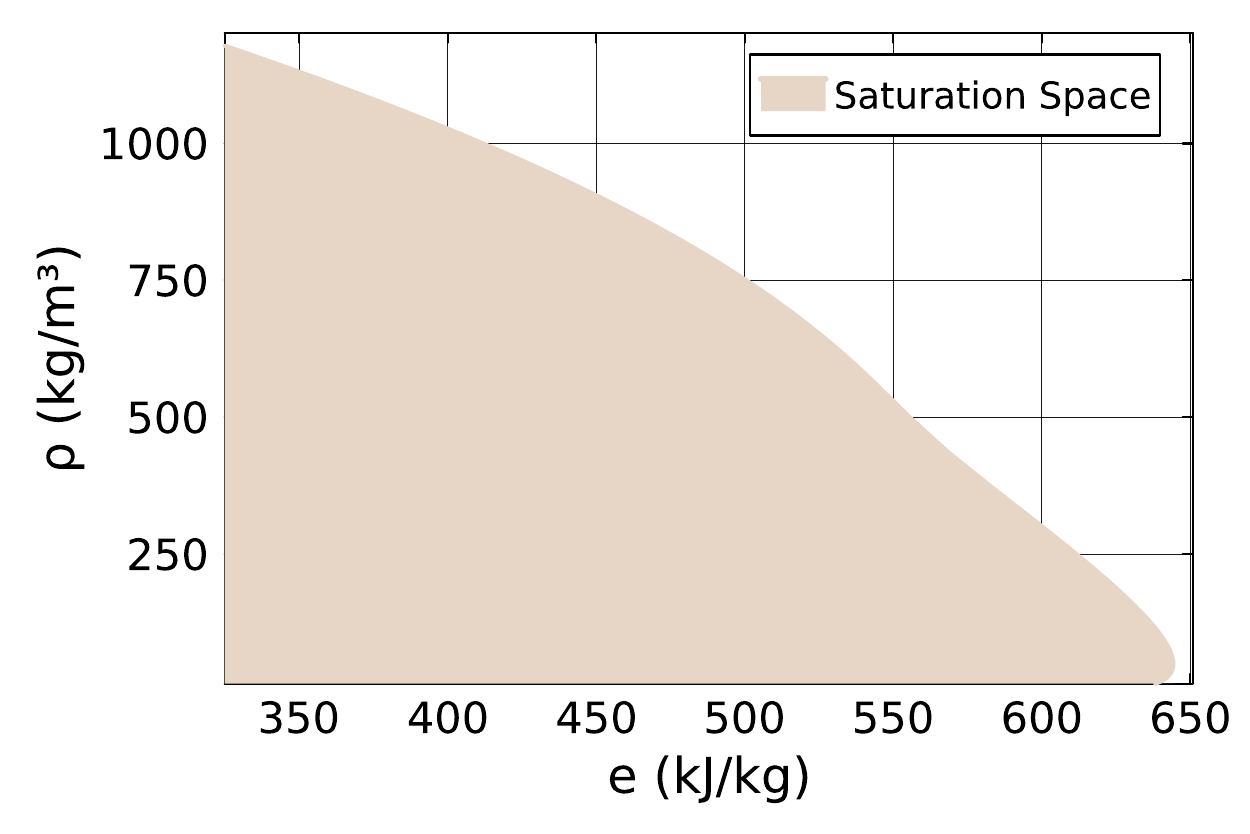}
         \caption{Saturation space in $\rho-e$ co-ordinates}
         \label{fig:SaturationCurve_RhoE}
     \end{subfigure}
        \caption{Saturation curve/space for \coo in $p-T$ and $\rho-e$ co-ordinates}
        \label{fig:SaturationCurve}
\end{figure}
Figures \ref{fig:SaturationCurve} (a)-(b) show the saturation region generated using the Span-Wagner EOS. Note that the saturation region is a curve in $p-T$ space whereas it corresponds to an area in $\rho-e$ space. A point ($(T,p)$ or $(e, \rho)$) is in the two-phase conditions if it lies on the saturation curve in $p-T$ space and inside the saturation area in $\rho-e$ space, otherwise it is in single phase. If the system is in single phase conditions, and for example $\rho=\Tilde{\rho}$ and $e=\Tilde{e}$ are given, we can solve 
\begin{equation}
\label{single-phase-eqm}
  e(\Tilde{\rho}, T) = \Tilde{e}, 
\end{equation}
to determine the temperature $T$. If the system is in two phase conditions, we need to solve a system of algebraic equations representing the vapour-liquid equilibrium to compute the various thermodynamic quantities. In literature this is also known as a flash problem \cite{michelsen_state_1999}. For a single component system like pure \coo, the two-phase system is in thermodynamic equilibrium when the pressure, temperature and Gibbs free energy ($G = h - Ts$) of both phases are equal. Depending upon which inputs are given, different types of flash routines are available\footnote{PH flash for known pressure and enthalpy, PS flash for known pressure and entropy etc. Please refer to \cite{michelsen_state_1999} for a detailed overview of various types of flashes}. Since we can compute $\rho$ and $e$ from the conservative variables $U$ at each time step of a transient simulation, the UV flash \footnote{U stands for internal energy, and V for specific volume; this is in contrast to our notation where $U$ denotes conserved variables and $V$ primitive variables} is a natural choice in our case. Given $\Tilde{\rho}$ and $\Tilde{e}$, the UV flash can be formulated as a system of four equations for the four unknowns $\alpha_g$, $\rho_g$, $\rho_l$ and $T$ as follows:
\begin{equation}
\label{eqm_eqns}
\begin{aligned}
    \alpha_g \rho_g+ (1-\alpha_g)\rho _l &= \Tilde{\rho}, \\
    \alpha_g \rho_g e(\rho_g, T) + (1-\alpha_g)\rho _{l}e(\rho_l, T) &= \Tilde{\rho} \Tilde{e},  \\
    p\left(\rho_g, T\right)&=p\left(\rho_l, T\right), \\
    G\left(\rho_g, T\right)&=G\left(\rho_l, T\right).
\end{aligned}
\end{equation}
We need to know in advance if the fluid is in single-phase or two-phase conditions in order to determine whether equation \eqref{single-phase-eqm} or system \eqref{eqm_eqns} has to be solved. Discrimination between the single-phase and two-phase regime is done as follows. For a given temperature $T$, we compute the saturation vapour and liquid densities, $\rho_g$, $\rho_l$. If $\rho_g < \rho < \rho_l$, then the fluid is two-phase, otherwise it is single phase.

\section{New vapour-liquid equilibrium equations} \label{sec:redcued_eqns}

\subsection{Reduced algebraic vapour-liquid equilibrium equation} \label{subsec:redcued_eqns}

The system of equations \eqref{eqm_eqns} can be simplified in the case of a pure component. Consider a pure component in two-phase flow conditions with a given mixture density, $\tRho$ and mixture internal energy, $\te$. Once we specify the temperature, all other thermodynamic quantities, e.g.\ $p$, $\rho_g$, $\rho_l$, $e_g$ and $e_l$, can be determined through the saturation relations \cite{span_new_1996}. This enables us to parameterise the various thermodynamic quantities as a function of only temperature; in other words: 
\begin{align} \label{sat_relations}
        p &= p(T),\\
    \rho_g &= \rho_g(T),\\
    \rho_l &= \rho_l(T),\\
    e_g &= e(\rho_g(T), T),\\
    e_l &= e(\rho_l(T), T).    
\end{align}
These saturation relations implicitly contain the information encoded by the pressure equilibrium and Gibbs free energy equilibrium, i.e.\ the last two equations of \eqref{eqm_eqns}. Thus, if saturation relations are known explicitly, one can directly apply them and avoid solving these two equations numerically. Saturation relations are normally provided as ancillary equations for phase density and saturation pressure in terms of temperature, e.g.\ Span-Wagner \cite{span_new_1996}. In principle, these saturation relations can be constructed from the EOS to arbitrary accuracy using a procedure outlined in Bell et al.\ \cite{bell_efficient_2021}, who used Chebyshev expansions to create saturation relations. In this work, we use the saturation relations of Span-Wagner \cite{span_new_1996}. Table \ref{table:Uncertainty_in_Correlations} summarises the uncertainties in the saturation correlations linked to the Span-Wagner EOS.

\begin{table}[h!]
\centering
\caption{Uncertainty in Span-Wagner saturation correlations\label{table:Uncertainty_in_Correlations}, from \cite{span_new_1996}.}
\begin{tabular}{llll}
\toprule
Temperature range (K)  & $\triangle p(\%)$    & $\triangle \rho_l(\%)$ & $\triangle \rho_g(\%)$ \\ 
\midrule
$T_t \leq T \leq 295$ & $\pm 0.012$ & $\pm 0.015$   & $\pm 0.025$   \\ 
$295 < T \leq 303$  & $\pm 0.012$ & $\pm 0.04$    & $\pm 0.08$    \\ 
$303 < T \leq T_c$ & $\pm 0.012$ & $\pm 1 $      & $\pm 1$       \\ 
\bottomrule
\end{tabular}
\end{table}

\FloatBarrier

Given the saturation relations, we can simplify system \eqref{eqm_eqns}. Recalling the definition of mixture density, we have
\begin{align}
\alpha_g \rho_g(T)  + (1-\alpha_g) \rho_l(T)  = \tRho \implies \alpha_g = \frac{\rho_l(T) - \tRho}{\rho_l(T) - \rho_g(T)}.   
\end{align}
Substituting the expression for $\alpha_g$ into the mixture internal energy equation,
\begin{equation}
\label{mix-int-energy}
\alpha_g \rho_g(T) e_g(T) + (1-\alpha_g) \rho_l(T) e_l(T) = \tRho \te ,
\end{equation}
we have 
\begin{align}\label{mix-int-energy2}
\frac{\rho_l(T) - \tRho }{\rho_l(T) - \rho_g(T)} \rho_g(T) e_g(T) + \frac{\tRho - \rho_g(T)}{\rho_l(T) - \rho_g(T)} \rho_l(T) e_l(T) = \tRho \te.
\end{align}
Defining 
\begin{align}
\fg(\rho, T) &:= \frac{\rho_l(T) - \rho}{\rho_l(T) - \rho_g(T)} \rho_g(T) e_g(T),\\
\fl(\rho, T) &:= \frac{\rho - \rho_g(T)}{\rho_l(T) - \rho_g(T)} \rho_l(T) e_l(T),   
\end{align}
equation \eqref{mix-int-energy2} can be rewritten as
\begin{align}
\label{int-energy-abstract}
\fg( \tRho, T) + \fl(\tRho, T) = \tRho \te.
\end{align}
For given $\tRho$ and $\te$, equation \eqref{int-energy-abstract} can be solved for $T$. In summary, by using the saturation relations we have reduced the system of four equations \eqref{eqm_eqns} to a single equation, namely equation \eqref{int-energy-abstract}. This is our first step in reducing the computational expense of the coupled problem of fluid flow and thermodynamics.

Equations \eqref{single-phase-eqm} and \eqref{int-energy-abstract} can be written in a succinct form as
\begin{align}
    &\f(\tRho, T) = \tRho \te, \label{reduced_vle}
\end{align}  
\begin{align}
    &\text{where} \quad \f(\tRho, T) :=\begin{cases}
			\quad \tRho e(\tRho, T), & \text{if single phase,}\\
            \quad \fg(\tRho, T) + \fl(\tRho, T), & \text{if two phase.}
		 \end{cases}
\end{align}
We shall refer to equation \eqref{reduced_vle} as \reducedDae, whereas equations \eqref{single-phase-eqm} and \eqref{eqm_eqns} will be referred to as \fullDae in the rest of this document.

\subsection{Reduced differential vapour-liquid equilibrium equation }\label{subsec:ode_derivation}

Equation \eqref{reduced_vle} is a non-linear equation in terms of the temperature, $T$. The common approach is to use a non-linear solver such as Newton-Raphson to solve this equation (see, for instance, \cite{giljarhus_solution_2012}, \cite{hammer_method_2013}). Here we propose an alternative approach which stems from the fact that the thermodynamic equilibrium (flash) problem is solved in conjunction with the time-dependent fluid flow conservation equations (for pipe or tank) as described in sections \ref{sec:HEM} and \ref{sec:tank_equations}. This means that the temperature is a function of time, $T(t)$, like the other quantities such as $\rho(t)$, $p(t)$, etc.

Our key insight is that we can derive an ODE for the evolution of the temperature by differentiating the reduced VLE, equation \eqref{reduced_vle}, with respect to $t$. Assuming that $\f(\rho,T)$ is differentiable in time, we get 
\begin{align}
    \EvaluatedAt{\frac{\partial \f}{\partial \rho}}{T} \frac{\rd\rho}{\rd t} + \EvaluatedAt{\frac{\partial \f}{\partial T}}{\rho} \frac{\rd T}{\rd t} = \frac{\rd (\rho e)}{\rd t},
\end{align}
where the subscripts $T$ and $\rho$ indicate that a differential is evaluated at constant $T$ or $\rho$, respectively. This equation can be rewritten into the following \textit{temperature evolution equation}:
\begin{align}    
   \boxed{
    \frac{\rd T}{\rd t} = \left[\frac{\rd(\rho e)}{\rd t} - \EvaluatedAt{\frac{\partial \f}{\partial \rho}}{T}\frac{\rd \rho}{\rd t}\right]/\EvaluatedAt{\frac{\partial \f}{\partial T}}{\rho} \label{temp-eqn}
    }.
\end{align}
We call this the \ode approach. In this approach, \textit{the temperature evolution in time is determined such that the thermodynamic equilibrium equation for the internal energy is always satisfied}, provided that it is satisfied by the initial conditions.

The time derivatives of mixture internal energy and density, $\frac{\rd \rho e}{\rd t}$ and $\frac{\rd \rho}{\rd t}$ respectively, are given by the fluid flow conservation equations for the pipe (equation \eqref{eqn:semiDiscrete2}) or tank (equation \eqref{tank_eqns}). 
The partial derivatives of $\f$ in equation \eqref{temp-eqn} are as follows.\\\\
\textbf{Single Phase Case:}
\begin{align}
    \EvaluatedAt{\frac{\partial \f}{\partial \rho}}{T} &= \rho \EvaluatedAt{\frac{\partial e}{\partial \rho}}{T} + e,\\    
    \EvaluatedAt{\frac{\partial \f}{\partial T}}{\rho} &= \rho \EvaluatedAt{\frac{\partial e}{\partial T}}{\rho}    
\end{align}
\textbf{Two Phase Case:}
\begin{align}
   \EvaluatedAt{\frac{\partial \f}{\partial \rho}}{T} &= \EvaluatedAt{\frac{\partial \fg}{\partial \rho}}{T}, \\
    \EvaluatedAt{\frac{\partial \f}{\partial T}}{\rho} &= \EvaluatedAt{\frac{\partial \fg}{\partial T}}{\rho} + \EvaluatedAt{\frac{\partial \fl}{\partial T}}{\rho}.
\end{align}

\begin{remark}
    $\EvaluatedAt{\frac{\partial \f}{\partial T}}{\rho}$
\end{remark}

\subsection{Summary of equations for VLE} \label{sec:summary_eqns}
We have so far discussed three ways of performing thermodynamic calculations in single and two phase conditions. Table \ref{tbl:summary_eqns} summarises the various alternatives.

\begin{table}[h]
\caption{The three approaches for incorporating thermodynamic constraints as investigated in this work. \label{tbl:summary_eqns}}
\begin{tabular}{@{}p{0.15\textwidth} p{0.42\textwidth} p{0.42\textwidth}@{}}
\toprule
Approach & Single Phase & Two Phase                                      
 \\ \midrule
\begin{tabular}{@{}p{0.15\textwidth} p{0.42\textwidth} p{0.42\textwidth}@{}} \fullDae \\ \\ equation \eqref{eqm_eqns} \end{tabular} & \(
        \begin{aligned}
            &e(\tRho, T) = \te \\\\\\\\\\
            & \text{Unknowns: } T                        
        \end{aligned}
        \)
& \begin{tabular}[c]{@{}l@{}} 
    \(
        \begin{aligned}
            &\alpha_g \rho_g+ (1-\alpha_g)\rho _l = \tRho \\
            &\alpha_g \rho_g e(\rho_g, T) + (1-\alpha_g)\rho _{l}e(\rho_l, T) = \tRho \te  \\
            &p\left(\rho_g, T\right)=p\left(\rho_l, T\right) \\
            &G\left(\rho_g, T\right)=G\left(\rho_l, T\right)\\\\
            & \text{Unknowns: } [\alpha_g, \rho_g, \rho_l, T]^T            
        \end{aligned}
    \)
\end{tabular} \\ \midrule
\begin{tabular}{@{}p{0.15\textwidth} p{0.42\textwidth} p{0.42\textwidth}@{}} \reducedDae \\ \\ equation \eqref{reduced_vle} \end{tabular} & \(
        \begin{aligned}
            &\psi(\tRho, T) =\tRho \te \\\\
            &\text{Unknowns: } T                        
        \end{aligned}
        \)
& \(
    \begin{aligned}
        &\fg(\tRho, T) + \fl(\tRho, T) = \tRho \te\\\\
        &\text{Unknowns: } T
    \end{aligned}
\)                                                           \\ \midrule
\begin{tabular}{@{}p{0.15\textwidth} p{0.42\textwidth} p{0.42\textwidth}@{}} \ode \\ \\ equation \eqref{temp-eqn} \end{tabular} & \(
\begin{aligned}
    &\frac{dT}{dt} =  \left[\frac{d(\rho e)}{dt} - \EvaluatedAt{\frac{\partial \f}{\partial \rho}}{T} \frac{d \rho}{dt}\right]/\EvaluatedAt{\frac{\partial \f}{\partial T}}{\rho}\\
    &\text{where } \f(\tRho, T) = \tRho e(\tRho, T)\\\\
    &\text{Unknowns: }  T
    \end{aligned}
\)
& \(
\begin{aligned}
    &\frac{dT}{dt} = \left[\frac{d(\rho e)}{dt} - \EvaluatedAt{\frac{\partial \f}{\partial \rho}}{T} \frac{d \rho}{dt}\right]/\EvaluatedAt{\frac{\partial \f}{\partial T}}{\rho}\\
    &\text{where } \f(\tRho, T) = \fg(\tRho, T) + \fl(\tRho, T)\\\\
    &\text{Unknowns: } T
    \end{aligned}
\)                                                          \\ \bottomrule
\end{tabular}
\end{table}

We started with the \fullEqns, system \eqref{eqm_eqns}. For two-phase conditions, we used the fact that the saturation line can be parameterized by the temperature, and reduced the system of equations to a single equation, equation \eqref{reduced_vle}. The main advantage of this approach is that the flash problem in both single phase and two phase conditions becomes a single equation with a single unknown quantity (namely temperature), which makes switching between single and two-phase conditions easier. The main assumption and the associated error with this approach lies in the accuracy of the applied saturation relations, as discussed in Table \ref{table:Uncertainty_in_Correlations}.

By differentiating the flash problem and incorporating the time derivatives of the mixture internal energy and density, we obtained a new ODE for the evolution of the temperature, equation \eqref{temp-eqn}. The temperature evolution equation can be used to perform time integration of the coupled fluid flow - thermodynamics without requiring a non-linear solver for the flash problem. The main assumption associated to this approach is that the solution is sufficiently smooth (differentiable) in time. 

\subsection{Coupling the ODE formulation to the tank and pipe equations}
For the tank problem, the full system of coupled flow and thermodynamics reads
\begin{align}
    \frac{\rd \rho}{\rd t} &= - \frac{\dot{m}(\rho, T)}{\Vol}, \label{tank:density} \\
    \frac{\rd T}{\rd t} &= \left[\frac{\rd(\rho e)}{\rd t} - \EvaluatedAt{\pd{\f}{\rho} }{T}\frac{\rd \rho}{\rd t}\right]/\EvaluatedAt{\pd{\f}{T}}{\rho} \label{tank:temperature}\\
    &= \frac{1}{\Vol}\left(\dot{Q}(T) - \dot{m}(\rho, T) h(\rho, T) + \EvaluatedAt{\pd{\f}{\rho}}{T} \dot{m}(\rho, T) \right)/\EvaluatedAt{\pd{\f}{T}}{\rho}.  \nonumber
\end{align}
The original system of two evolution equations for the fluid with thermodynamic constraints has thus been reduced to only two evolution equations. This ODE representation is computationally attractive when combined with explicit ODE solvers, such as the forward Euler method, so that the need to solve a nonlinear equation is entirely circumvented (both for the flow equations and the thermodynamics). 

In a similar fashion, the ODE system for the pipe problem can be derived. One important difference between the pipe and the tank is that the fluid flow equations and thermodynamic constraint are now written for each grid point $i$. Another difference is that the pipe equations feature the time evolution of $\frac{\rd (\rho E)}{\rd t}$, whereas the tank features $\frac{\rd (\rho e)}{\rd t}$.  
Therefore, we need to derive an equation for the evolution of the internal energy.

The time derivatives of conservative variables are available through equation \eqref{eqn:semiDiscrete2}. Writing \eqref{eqn:semiDiscrete2} for each conservative variable in a particular grid point $i$, we get
\begin{align}
     \frac{\rd \rho}{\rd t}   &= \fOne(U,T), \label{pipe:density}\\
     \frac{\rd (\rho u)}{\rd t} &= \fTwo(U,T), \label{pipe:momentum}\\
     \frac{\rd (\rho E)}{\rd t} &= \fThree(U, T). \label{pipe:energy}
\end{align}
where $\fOne$, $\fTwo$, $\fThree$ are discretised flux divergence operators for mass, momentum and energy  respectively.
Compared to \eqref{eqn:semiDiscrete2}, we have added $T$ as argument in $\fOne$, $\fTwo$, $\fThree$, since the pressure follows from thermodynamics relations of the form $p(U,T)$, see equation \eqref{p-hemholtz}. Using equations \eqref{pipe:density} and \eqref{pipe:momentum}, we can get the evolution equation for the (specific) kinetic energy:
\begin{equation}
    \dd{(\rho u^2/2)}{t} =\dd{((\rho u)^2/2 \rho)}{t} = u \dd{(\rho u)}{ t} - \frac{1}{2} u^2 \dd{\rho}{t} \label{pipe:kinetic-energy}
\end{equation}
Now, using equations \eqref{pipe:energy} and \eqref{pipe:kinetic-energy}, we can get an equation for the evolution of the internal energy:
\begin{align}
    \dd{\rho e}{t} &= \dd{(\rho E)}{t} - \dd{(\rho u^2/2)}{t} \\
                    &= \dd{(\rho E)}{t} - u \dd{(\rho u)}{ t} + \frac{1}{2} u^2 \dd{\rho}{t}.
\end{align}
This equation can be substituted in equation \eqref{temp-eqn} to yield the evolution equation for the temperature in grid point $i$:
\begin{align} \label{temp-eqn-euler}
    \dd{T}{t} &= \left(\dd{(\rho E)}{t} - u \dd{(\rho u)}{ t} + \frac{1}{2} u^2 \dd{\rho}{t} - \EvaluatedAt{\pd{\f}{\rho}}{T} \dd{\rho}{t}\right) / \EvaluatedAt{\pd{\f}{T}}{\rho}. 
\end{align}

 Using \eqref{pipe:density}, \eqref{pipe:momentum} and \eqref{pipe:energy} in \eqref{temp-eqn-euler}, we have
\begin{equation} \label{pipe:temperature}
\dd{T}{t} = \left(\fThree(U,T) - u \fTwo(U,T)  + \frac{1}{2} u^2 \fOne(U, T) - \EvaluatedAt{\pd{\f}{\rho}}{T} \fOne(U, T)\right) / \EvaluatedAt{\pd{\f}{T}}{\rho}.
\end{equation}
The system of equations \eqref{pipe:density}, \eqref{pipe:momentum}, \eqref{pipe:energy} and \eqref{pipe:temperature} constitutes a complete set of governing equations that can be used to advance the pipeline simulation to the subsequent time level. The thermodynamic constraint (vapour-liquid equilibrium) is encoded into these equations, albeit in a time-differentiated fashion. Upon time integration, errors can accumulate, which can cause errors in the satisfaction of equation \eqref{reduced_vle}.

To avoid this issue, we propose to solve instead the following set of equations: equations \eqref{pipe:density}, \eqref{pipe:momentum}, \eqref{pipe:temperature} and \eqref{reduced_vle}, written as
\[
 \rho E = \f(\rho, T)+ \frac{1}{2} \rho u^2 \label{pipe:total_energy_postprocessed}.
\]
For the time-continuous system, these four equations are equivalent to using the four equations  \eqref{pipe:density}, \eqref{pipe:momentum}, \eqref{pipe:energy} and \eqref{pipe:temperature} as the energy conservation equation \eqref{pipe:energy} can be derived from the temperature equation and the thermodynamic constraint. Upon discretizing in time with forward Euler, this equivalence is lost and a first order temporal error is introduced in the energy equation. 

In summary, we have traded exact energy conservation with an inexact thermodynamic constraint for an exact thermodynamic constraint with inexact energy conservation. In section \ref{sec:ode_approach}, the associated energy conservation error will be numerically investigated. 



\FloatBarrier

\section{Time integration methods}\label{sec:numerical_method}

In the previous section, three different approaches for simulating fluid flows coupled with thermodynamic constraints were introduced. In this section, we discuss the time integration methods applied for each approach.

\subsection{Algebraic approaches: DAE interpretation}\label{sec:DAE_interpretation}
We first consider the algebraic approaches \fullDae and \reducedDae. The combination of the semi-discrete pipe flow equations \eqref{eqn:semiDiscrete} (or tank equations \eqref{tank_eqns}) and thermodynamics (see Table \ref{tbl:summary_eqns}) forms a system of $3N$ (pipe) or 2 (tank) ODEs with $N_{F}$ algebraic constraints. This constitutes a system of differential-algebraic equations (DAE). For both the pipe and the tank problem, we can write the DAE system in the so-called semi-explicit form \cite{ascher_computer_1998}:
\begin{align}
    \frac{\rd U}{\rd t} &= f(U,V), \label{dae_eqns_f} \\
    0 &= g(U, V). \label{dae_eqns_g}
\end{align}
where $U$ and $V$ are the vectors of conservative and non-conservative variables, respectively. For \fullDae, if $N_{TP}$ cells exhibit two-phase conditions, then $N_{F} = 4N_{TP} + (N - N_{TP}) = 3 N_{TP} + N$ for pipes. For the tank, $N_{F} = 1$ for single-phase conditions and $N_{F} = 4$ for two-phase conditions. For \reducedDae, two-phase and single-phase conditions both require a single algebraic equation, so $N_F = N$ for the pipe and $N_{F}=1$ for the tank.  

For the pipe flow equations, 
\begin{align*}
  &U = [U_{1}, \ldots, U_{N}]^T, \qquad U_{i} = [\rho_i, (\rho u)_i, (\rho E)_i]^T,\\
  &V = [V_{1}, \ldots, V_{N}]^T
\end{align*}
where $V_i = [\alpha_{g,i}, \rho_{g,i}, \rho_{l,i}, T_i]^T$ for \fullDae and $V_i = T_i$ for \reducedDae. Furthermore, $f$ represents the spatial discretization of the fluid flow equations,
\begin{equation}
    f_i(U, V) := -\frac{1}{\triangle x} (\Hat{\EulerFlux}_{i+\frac{1}{2}}(U_{i+1},U_{i},V_{i+1}, V_{i}) - \Hat{\EulerFlux}_{i-\frac{1}{2}}(U_{i},U_{i-1},V_{i}, V_{i-1})),
\end{equation}
and $g$ represents either the full or reduced VLE equation(s) for the $i^{th}$ cell. For example, the first equation of \eqref{eqm_eqns} for \fullDae for grid cell $i$ in two-phase conditions reads:
\begin{equation}
    g_{i,1} := \alpha_{g, i} \rho _{g, i}+ (1-\alpha_{g, i})\rho _{l, i} - \rho_i  = 0.
\end{equation} 

For the tank equations, 
\begin{align*}
  U = [\rho, \rho e]^T, 
\end{align*}
and $V$ is the same as for the pipe and $f$ is defined as per equation \eqref{tank_eqn_dae_rhs}. 

To solve the DAE system \eqref{dae_eqns_f} in time we use half-explicit forward Euler method \cite{hairer_numerical_1989}, which constitutes an explicit update for the evolution equation \eqref{dae_eqns_f}, followed by a nonlinear equation solve for the constraint \eqref{dae_eqns_g}. 
We first solve
\begin{align} \label{eq:forward_euler_with_constraint}
    U^{n+ 1} &= U^{n} + \Delta t f(U^n, V^n),
\end{align}
for $U^{n+1}$, followed by solving the constraint equation
\begin{equation}
    0 = g(U^{n+1}, V^{n+1}),
\end{equation}
for $V^{n+1}$ with the Newton-Raphson with Line search. 


\subsection{ODE approach} \label{sec:ode_approach}
The system of equations \eqref{pipe:density}, \eqref{pipe:momentum}, and \eqref{pipe:temperature} for the pipe, and \eqref{tank:density} and \eqref{tank:temperature} for the tank, collectively form a system of ODEs. Specifically, the pipe system comprises \( 3N \) ODEs, while the tank system consists of \( 2 \) ODEs. 
Once the temperature \( T \) and density \( \rho \) are available, the internal energy \( e \) for both the pipe and the tank can be computed using \eqref{reduced_vle}. 
For the pipe, the total energy \( E \) is computed by incorporating the kinetic energy term, which will be detailed shortly.

For both the pipe and tank problems, the ODE system can be expressed as follows:
\begin{align}
\frac{dU}{dt} &= f(U), \label{ode_eqns_f}
\end{align}
where $U$ and $f$ need to be defined specifically for the pipe and tank problems, and are differing here from the definitions used in the DAE approach. For the pipe flow equations they take the form,
\begin{align*}
&U = [U_{1}, \ldots, U_{N}]^T, \qquad U_{i} = [\rho_i, (\rho u)_i, (T)_i]^T,\\
&E_i = \frac{\f(\rho_i, T_i)}{\rho_i} + \frac{1}{2} u_i^2
\end{align*}
\begin{equation*}
f_i(U) = -\frac{1}{\Delta x} (\hat{\EulerFlux}_{i+\frac{1}{2}}(U_{i+1},U_{i}) - \hat{\EulerFlux}_{i-\frac{1}{2}}(U_{i},U_{i-1})),
\end{equation*}
whereas for the tank equations they take the form,
\begin{align*}
&U = [\rho, T]^T, \\ \nonumber
&f(U) = \frac{1}{\Vol}\begin{bmatrix}
    - \dot{m}(\rho, T) \\ \left(\dot{Q}(T) - \dot{m}(\rho, T) h(\rho, T) + \EvaluatedAt{\pd{\f}{\rho}}{T} \dot{m}(\rho, T) \right)/\EvaluatedAt{\pd{\f}{T}}{\rho}
\end{bmatrix}.
\end{align*}

We use forward Euler to integrate equation \eqref{ode_eqns_f} in time.
\begin{align}
U^{n+ 1} &= U^{n} + \Delta t f(U^n).
\end{align}
Since we are using a numerical integrator rather than directly solving the algebraic constraints to compute the temperature in this approach, the numerical integration error(order of \odtt\ for forward euler) in temperature will propagate as an error in the computed energy. We will discuss this in the numerical experiments in Section \ref{sec:results}.
In addition, the equation \eqref{temp-eqn} contains the term $\pd{\f}{T}$ in the denominator. If this term approaches zero, it could cause numerical issues. However, we did not encounter this difficulty in the conducted numerical experiments.

\begin{remark} 
For improved accuracy, we use automatic differentiation to evaluate the partial derivatives $\pd{\f}{\rho}$ and $\pd{\f}{T}$ instead of finite differences.
\end{remark}

 \subsection{Transition from single-phase to two-phase}\label{sec:transition}
The transition from single phase to two phase during time integration constitutes a discontinuity in the definition of the ODE (or DAE) system. This discontinuity is typically characterized by what is referred to in the literature as a switching function \cite{ascher_computer_1998}. One approach to detect the switching point is through the utilization of an event location algorithm. In our case, the switching function returns a boolean value indicating whether the system is in a two-phase state or not. Our switching function is described as follows:

\begin{enumerate}
    \item For a given temperature \( T \), we compute the saturation vapour and liquid densities, \( \rho_g \) and \( \rho_l \).
    \item We discretize time into discrete intervals using uniform time steps \( \Delta t = t_{i+1} - t_{i} \), denoted as \( \{t_1, t_2, ..., t_i, t_{i+1}, ..., t_{N_t}\} \). To advance the solution from \( t_i \) to \( t_{i+1} \), we employ the forward Euler method. Before progressing to the subsequent time level, we verify if \( \rho_g < \rho < \rho_l \). If this condition holds, the fluid is in a two-phase state; otherwise, it is classified as single phase. It is possible that phase transitions between two-phase and single-phase states might occur between time steps due to solution updates. While this necessitates an iterative process to reconcile temperature, pressure, and phase state at each time step, the current study adopts a simplified approach by proceeding to the next time step without explicitly addressing these complexities.

\end{enumerate} 

\FloatBarrier

\section{Numerical experiments}\label{sec:results}


\subsection{Tank depressurization}\label{sec:tank_depressurization}
In this section, we conduct a depressurization simulation of a tank containing \coo, as described by equation \eqref{tank_eqns}. The objective is to demonstrate the performance of the two proposed approaches (\reducedDae, \ode) against \fullDae on the test case outlined by \Hammer \cite{hammer_method_2014}. Despite its apparent simplicity, this ODE system (equations \eqref{tank_eqns}) serves as an effective test case for the thermodynamic flash problem as it simulates the transition of the fluid from its initial liquid phase to a two-phase state (boiling of \coo), and finally, to a completely gaseous phase.

\subsubsection{Problem setup}

The initial conditions and model coefficients are summarized in Table \ref{table:Tank-Params}. The initial conditions are defined in terms of pressure ($p$) and temperature ($T$). To compute the initial density ($\rho$) and internal energy ($e$), we solve the equation $p_0 = p(\rho, T_{0})$ for $\rho$, where $p_0$ and $T_0$ denote the initial pressure and temperature in the tank, respectively. Once $\rho$ is determined, $e$ is computed using equation \eqref{e-hemholtz}. Additionally, it is important to note that \Hammer utilized a test case that eventually leads to three-phase (liquid/gas/solid) conditions as the simulation progresses. However, as the framework outlined in this paper is currently developed only for single and two-phase scenarios, we terminated the simulation before it entered the three-phase regime.

\begin{table}[h!]
\caption{Tank simulation parameters\label{table:Tank-Params}}
\begin{tabular}{llllllllll}
\toprule
 $p_{0}$(bar) & $T_0(K)$ & $p_{\text{amb}}$(bar) & $T_{\text{amb}}(K)$ & $\eta A$ (W/K) & $K_{v}(m^2)$ & $\Vol(m^3)$  & $\text{Simulation time($h$)}$                   \\ 
\midrule
  100           & 300          & \quad 10               & 293.15          & \quad 1                    & $5\cdot 10^{-7}$ & $\pi \cdot 10^{-2} $ & \quad 0.6 \\

\bottomrule
\end{tabular}
\end{table}
    
\subsubsection{Validation with Hammer}
\begin{figure}
     \centering
     \begin{subfigure}[b]{0.48\textwidth}
         \centering
         \includegraphics[width=\textwidth]{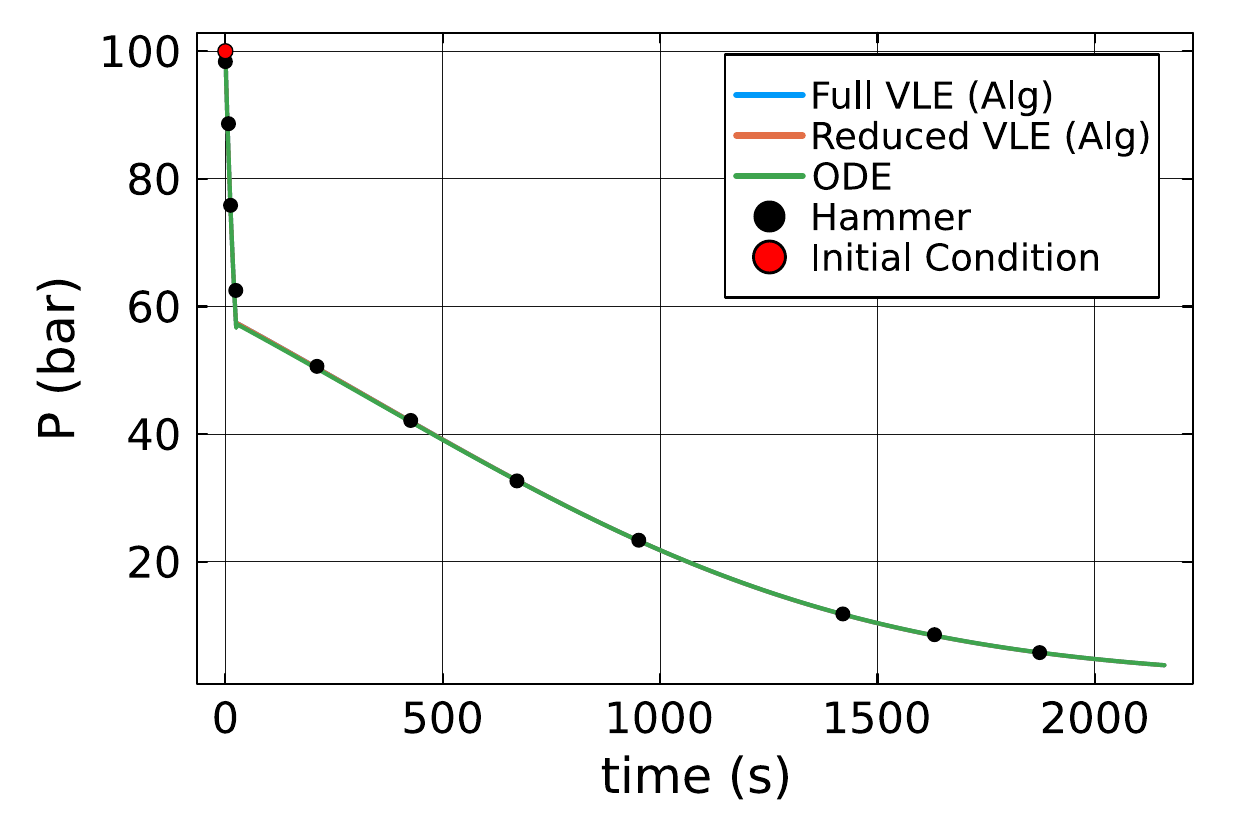}
         \caption{Pressure variation with time}
         \label{fig:Pressure_Tank_Hammer_SW}
         \vspace{6pt}
     \end{subfigure}
     \begin{subfigure}[b]{0.48\textwidth}
         \centering
         \includegraphics[width=\textwidth]{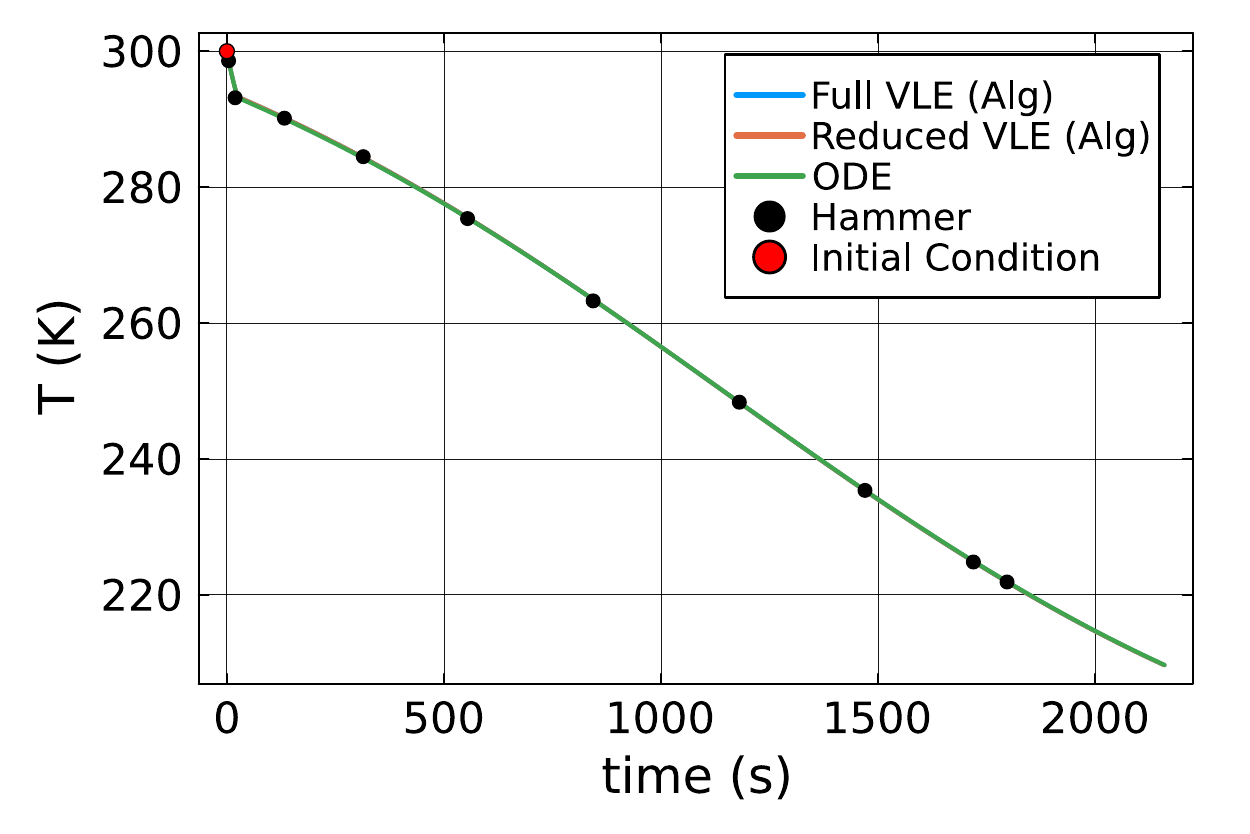}
         \caption{Temperature variation with time}
         \label{fig:Temperature_Tank_Hammer_SW}
         \vspace{6pt}
     \end{subfigure}
        \caption{Comparison of tank depressurization results with \Hammer $\Delta t = 1$\ s.}
        \label{fig:Validation_with_Hammer}
\end{figure}

First, we establish the validity of our simulation results by comparing them with those by \Hammer as shown in Figure \ref{fig:Validation_with_Hammer}. All three approaches demonstrate excellent agreement with those reported in \Hammer 

\begin{figure}
     \centering
     \begin{subfigure}[b]{0.48\textwidth}
         \centering
         \includegraphics[width=\textwidth]{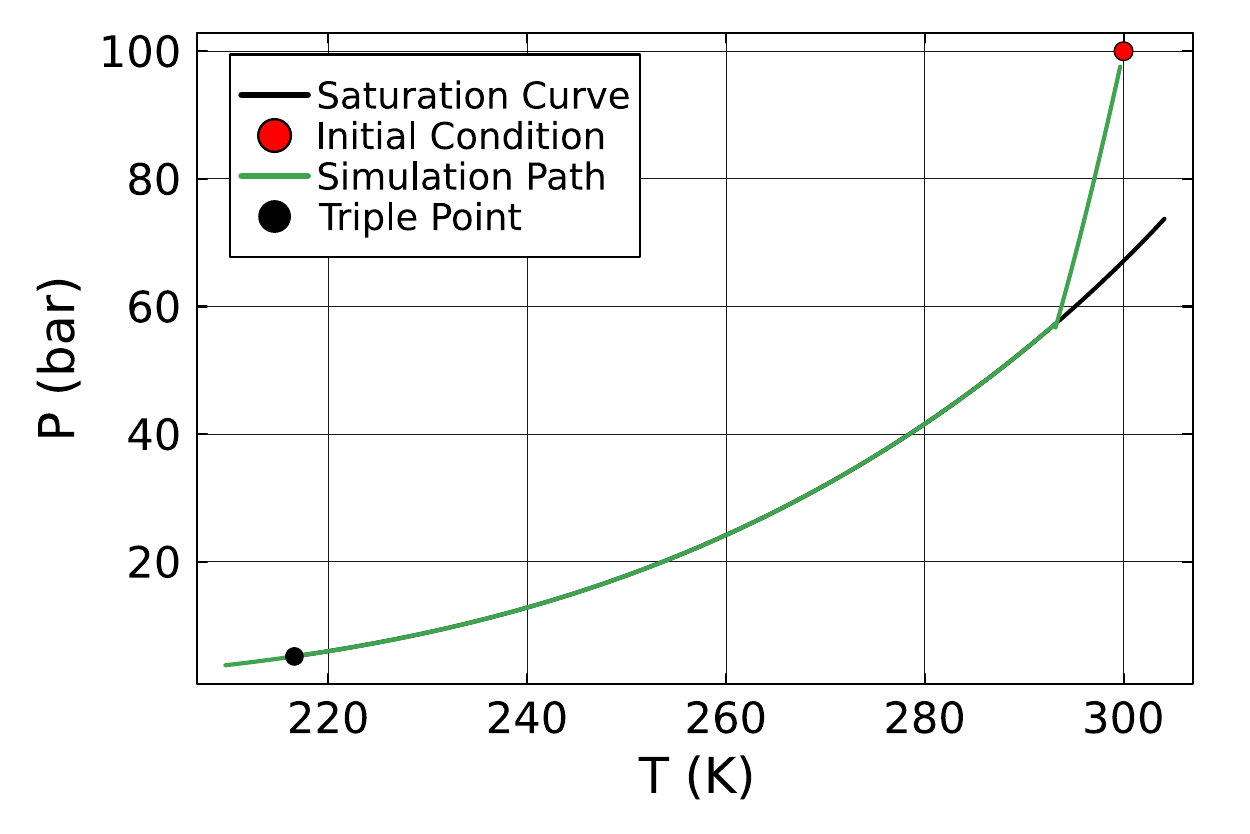}
         \caption{Simulation path in $p-T$ space}
         \label{fig:SimulationPathInPT}
         \vspace{6pt}
     \end{subfigure}
     \hfill
     \begin{subfigure}[b]{0.48\textwidth}
         \centering
         \includegraphics[width=\textwidth]{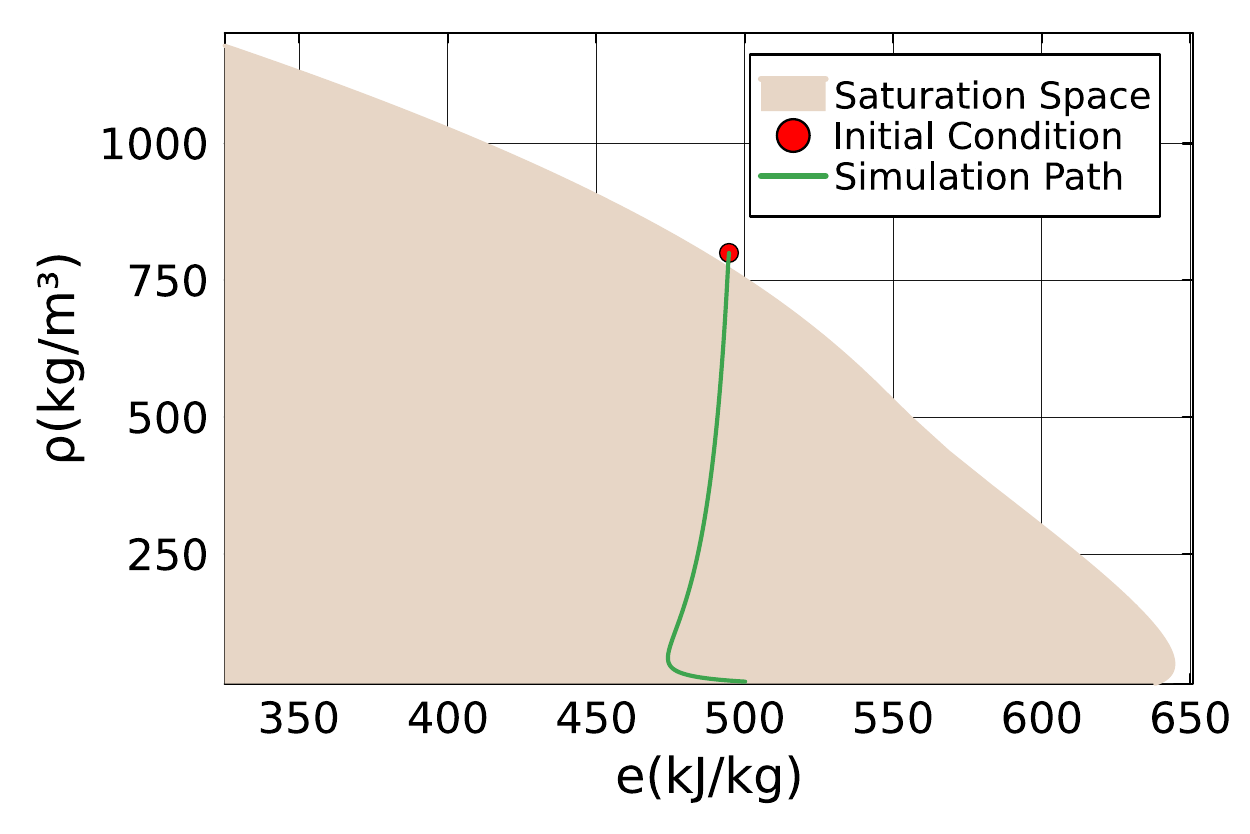}
         \caption{Simulation path in $\rho-e$ space}
         \label{fig:SimulationPathInRhoE}
         \vspace{6pt}
     \end{subfigure}
        \caption{Simulation trajectory for the \Hammer test case, $\Delta t = 1$\ s.}
        \label{fig:Simulation_Trajectory}
\end{figure}

Figure \ref{fig:Simulation_Trajectory} illustrates the simulation trajectory in $p-T$ and $\rho-e$ spaces. In the $\rho-e$ space, points within the colored region denote the two-phase regime, while those outside indicate single-phase conditions. Initially, a rapid pressure drop (from 100 bar to 57 bar) occurs until the saturation curve is reached, marking the transition of liquid \coo into the two-phase region within approximately 26 seconds. Subsequently, the pressure and temperature decrease along the saturation curve, reaching a lowest temperature of \( 209.15 \, \text{K} \, (-64 ^\circ \text{C}) \) at \( t = 0.6 \, \text{h} \). At this point, \coo transitions into a three-phase state and the simulation is terminated. Notably, during the depressurization process much lower temperatures are reached than those given by either the initial conditions or the ambient conditions. This underscores the potential low temperature risk associated with Joule-Thompson cooling of \coo during a depressurization, especially during the transitions between single-phase and two-phase states.

\subsubsection{Convergence study}
 Next, we proceed to analyze the two errors that were introduced with the \reducedDae and \ode approaches: (i) the approximation error from introducing the saturation relations to replace the full flash problem, and (ii) possible constraint-drift error arising from differentiating the constraint in time. The first error (`approximation error') is computed as:
\begin{align}  
    \varepsilon_{p,\text{full-reduced}}(t) &=  \frac{|p_{\text{full-vle}}(t) - p_{\text{reduced-vle}}(t)|}{|p_{\text{full-vle}}(t)|}, \label{eq:approximation_error} \\
    \varepsilon_{T,\text{full-reduced}}(t) &=  \frac{|T_{\text{full-vle}}(t) - T_{\text{reduced-vle}}(t)|}{|T_{\text{full-vle}}(t)|}. \nonumber
\end{align}
The second error (`constraint error') is computed as
\begin{align} \label{eq:constraint_error}
    \varepsilon_{p,\text{vle-ode}}(t) =  \frac{|p_{\text{reduced-vle}}(t) - p_{\text{ode}}(t) |}{|p_{\text{reduced-vle}}(t)|}, \\ 
    \varepsilon_{T,\text{vle-ode}}(t) =  \frac{|T_{\text{reduced-vle}}(t) - T_{\text{ode}}(t) |}{ |T_{\text{reduced-vle}}(t)|}. \nonumber
\end{align}

\begin{figure}
        \centering
        \begin{subfigure}[b]{0.49\textwidth}
        \includegraphics[width=\textwidth]{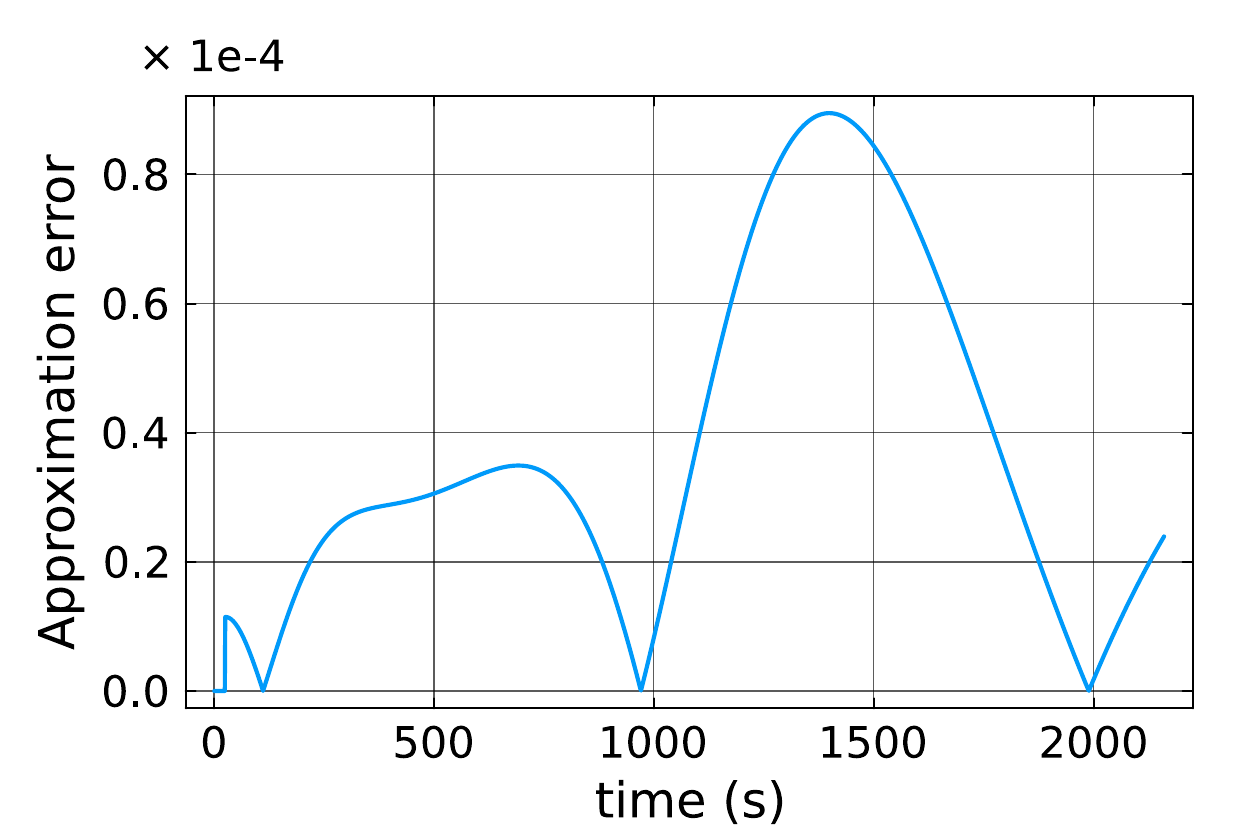}
        \caption{Error in pressure}
        \end{subfigure}
        \begin{subfigure}[b]{0.49\textwidth}
         \includegraphics[width=\textwidth]{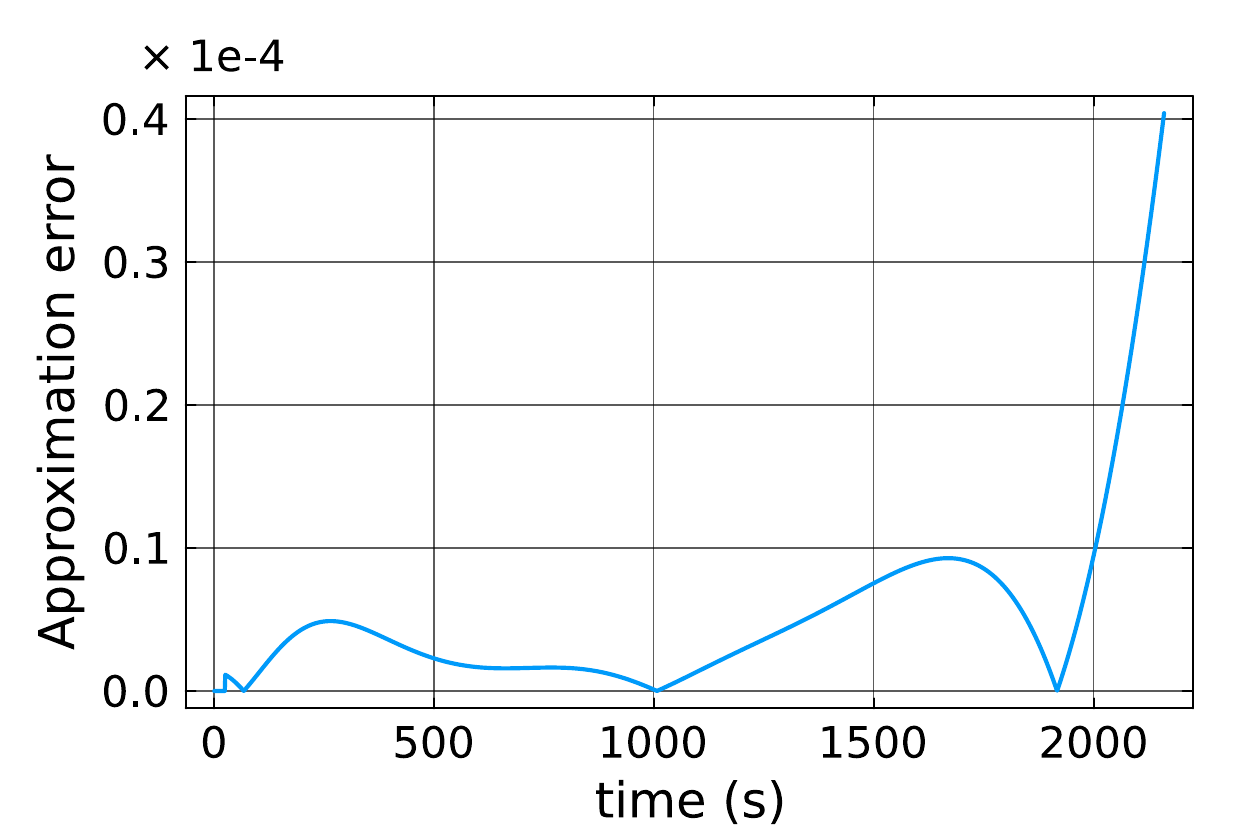}
         \caption{Error in temperature}
     \end{subfigure}
     \caption{Approximation error from \fullDae to \reducedDae, \dt{1}.}
     \label{fig:ApproximationErrorsFullToReduced}
\end{figure}

\begin{figure}
        \centering
        \begin{subfigure}[b]{0.48\textwidth}
        \includegraphics[width=\textwidth]{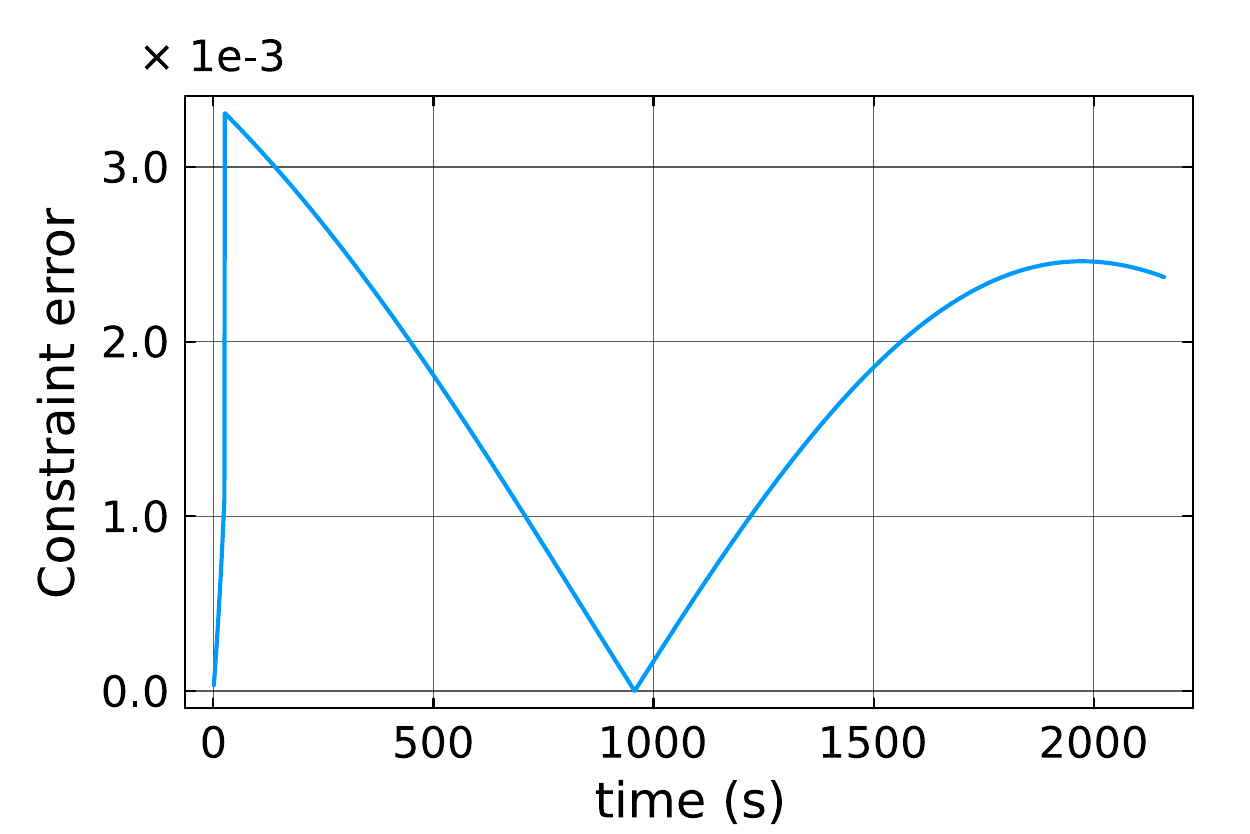}
        \caption{Error in pressure}
        \label{fig:ConstraintErrorInPressure}
        \end{subfigure}
        \begin{subfigure}[b]{0.48\textwidth}
        \includegraphics[width=\textwidth]{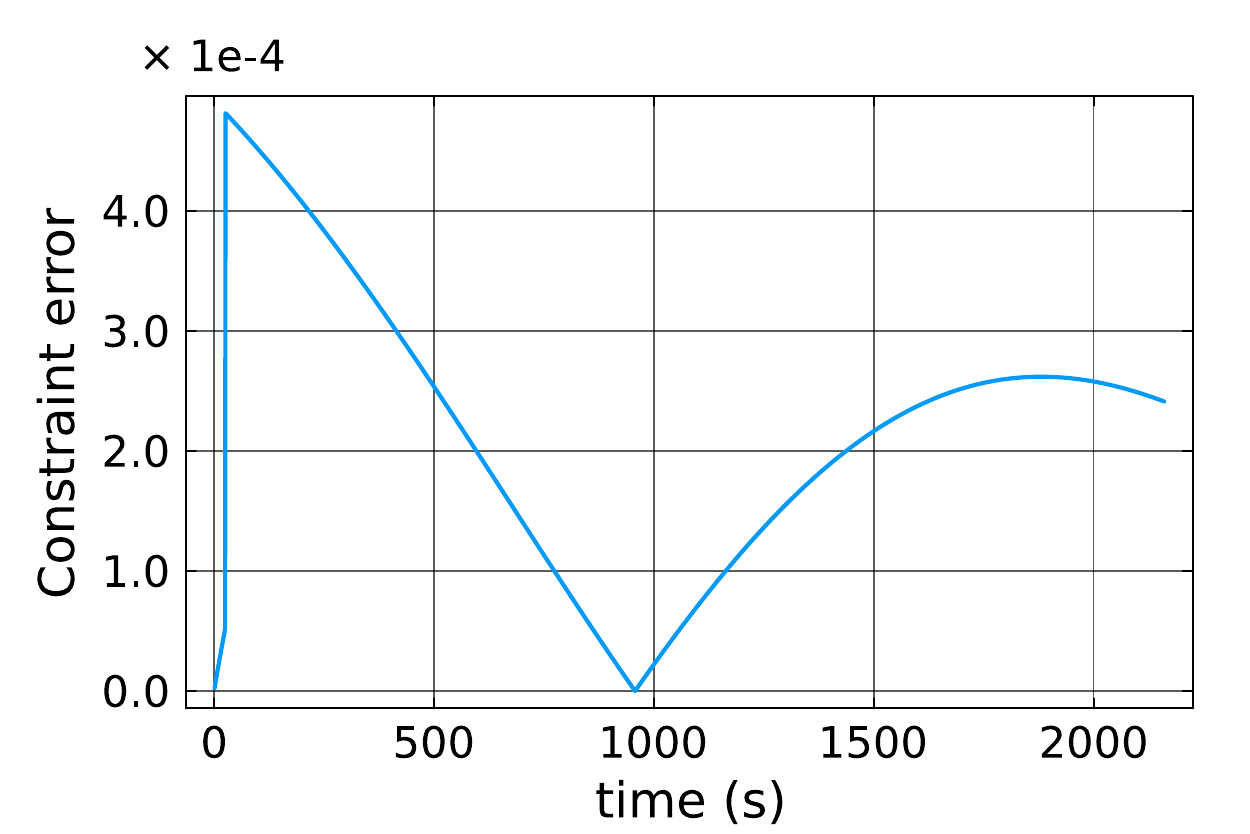}
        \caption{Error in temperature}
        \label{fig:ConstraintErrorInTemperature}
     \end{subfigure}
     \caption{Constraint error from \reducedDae to \ode, \dt{1}.}
     \label{fig:ConstraintErrorFromDaeToOde}
\end{figure}

Figure \ref{fig:ApproximationErrorsFullToReduced} shows the approximation errors \eqref{eq:approximation_error} incurred when going from the \fullDae approach to the \reducedDae approach. For example, the uncertainty in the pressure in the saturation relations is $\pm 0.012 \%$ (see Table \ref{table:Uncertainty_in_Correlations}) and the maximum error throughout the simulation stays well below this value (it is less than $0.01\%$). Figure \ref{fig:ConstraintErrorFromDaeToOde} illustrates the constraint errors in pressure and temperature, as defined by equation \eqref{eq:constraint_error}, introduced by transitioning from the \reducedDae to the \ode formulation for $\Delta t = 1$~s. Notably, the maximum errors are observed at $t \approx 26$ s, coinciding with the phase transition point. At this critical juncture, the pressure error reaches approximately $0.33\%$, while the temperature error remains around $0.048\%$. This underscores the necessity for a sufficiently small time step to accurately capture the rapid transients occurring around the phase transition point (here, $t \approx 26$ s). 

\begin{figure}[htbp]
    \centering
    \begin{subfigure}[b]{0.49\textwidth}
        \includegraphics[width=\textwidth]{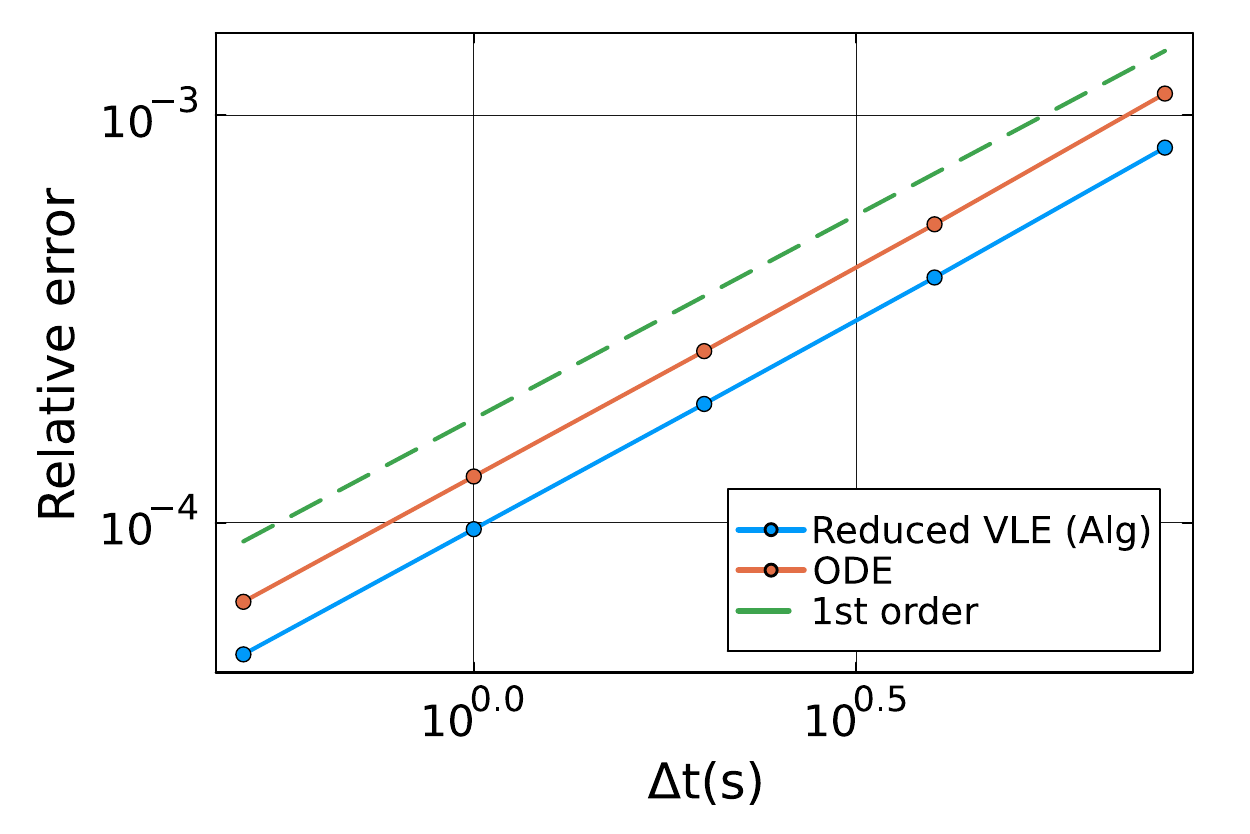}
        \caption{ Single-phase regime}
        \label{fig:TankHammerConvergenceSinglePhase}
    \end{subfigure}
    \hfill
    \begin{subfigure}[b]{0.49\textwidth}
        \includegraphics[width=\textwidth]{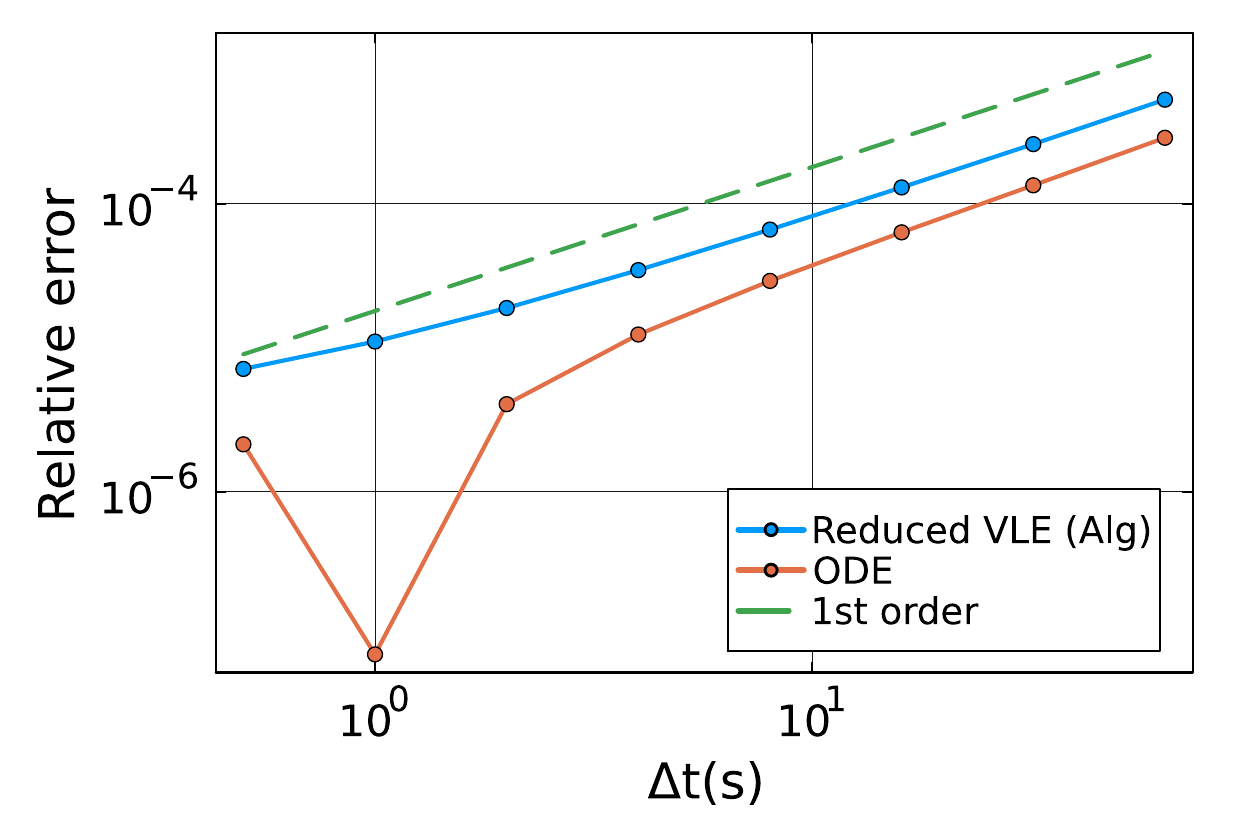}
        \caption{ Two-phase regime}
        \label{fig:TankHammerConvergenceTwoPhase}
    \end{subfigure}
    \caption{Relative temperature error for tank depressurization simulations with $\Delta t_\text{ref} = 1 \times 10^{-4}$\ s.}
    \label{fig:TankConvergence}
\end{figure}

Figures \ref{fig:TankHammerConvergenceSinglePhase} and \ref{fig:TankHammerConvergenceTwoPhase} depict the convergence analysis of our time integration method under distinct simulation scenarios. In Figure \ref{fig:TankHammerConvergenceSinglePhase}, the simulation remains within a single-phase regime throughout its entirety, and results are monitored at $t=16$ s. Conversely, in Figure \ref{fig:TankHammerConvergenceTwoPhase}, the simulation commences and persists in a two-phase state, with results monitored at $t=128$\ s. The errors in temperature are computed as $\varepsilon_{T} = (T - T_{\text{ref}})/T_{\text{ref}}$. For both approaches, the reference solution is obtained by solving with the \fullDae approach with \(\Delta t = 1 \times 10^{-4} \, \text{s}\). The observed convergence behaviour closely aligns with the expected outcomes for each method (first order convergence corresponding to the first order forward Euler method). However, for smaller \(\Delta t\) values (approximately 2 s) in the two-phase scenario, this first-order convergence degrades, indicating that residual errors persist due to the use of saturation relations. 
\subsubsection{Computational efficiency}
\begin{figure}[t]
    \centering
    \includegraphics[width=\textwidth]{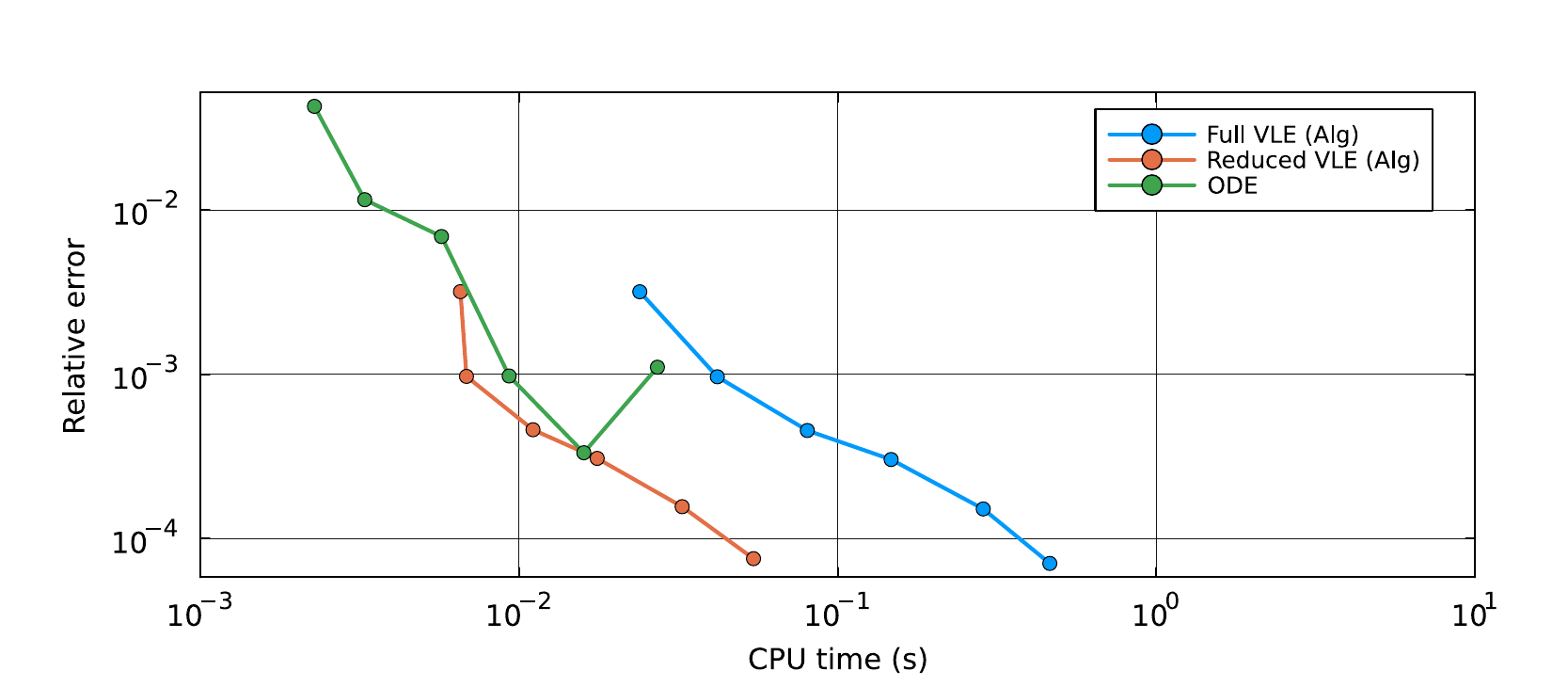}
    \caption{Error vs CPU time for all approaches.}
    \label{fig:TankHammerErrorVsTimeTakenRev2}
\end{figure}

In this subsection, we discuss the computational gains of our two new reduced approaches. Quantifying an exact speedup is challenging due to its dependence on the time step. However, Figure \ref{fig:TankHammerErrorVsTimeTakenRev2} illustrates the relative error in temperature plotted against CPU time for all three approaches. The reference solution for all approaches corresponds to the solution computed using \fullDae with \dt{1 \times 10^{-4}}. Notably, the reduced approaches require an order of magnitude less CPU time compared to the \fullDae approach. However, it is important to note that once the relative accuracy of the order of $10^{-3}$ is achieved, the performance gain in the ODE approach is lost. This can be attributed to the crossing of saturation boundary in the simulation which acts like a discontinuity, thus negating the advantages of the ODE approach beyond this accuracy threshold.


 
In conclusion, the reduced VLE approaches (both algebraic and ODE) demonstrate comparable performance, achieving significant speedup for relative accuracy of the order of $10^{-3}$. This indicates their suitability for practical applications where reduced computational overhead is necessary without significant loss of accuracy.


\FloatBarrier

\subsection{Pipeline depressurization}\label{sec:pipe_depressurization}
In this section, we consider a more difficult test case: the depressurization of a pipeline instead of a tank.
\subsubsection{Problem setup} \label{sec:pipe:problem_setup}
 We follow again \Hammer \ \cite{hammer_method_2014}, who describe the depressurization of a pipe that results from opening the right end of a pressurized pipe into the ambient. In \Hammer\ outflow boundary conditions are specified at the right end of a pipe of length $L=100\ m$. However, in our simulation, we avoid the use of outflow boundary conditions by doubling the pipe length to $L=200\ m$ and modelling it as a shock-tube problem. The rationale behind this is that the flow at the outlet will be choked so that information will not propagate upstream. The shock-tube has as initial condition a membrane in the middle, separating it into a left (pressurized) and right (ambient) section. The simulation is stopped before the fastest waves have reached the left or right end of the pipe, so that the solution at both ends ($x=0$ and $x=200\ m$) of the pipe will correspond to the initial conditions. It is noteworthy that the simulation configurations employed in \Hammer\ (characterized by outflow boundary conditions) and the present study (focusing on the shock-tube problem) are slightly different. Additionally, \Hammer\ utilizes the MUSTA scheme in conjunction with a strong-stability-preserving Runge-Kutta method. These methodological differences have the potential to induce deviations in the observed results, as will be subsequently demonstrated.

\begin{figure}[t]
    \centering
    \includegraphics[width=0.5\textwidth]{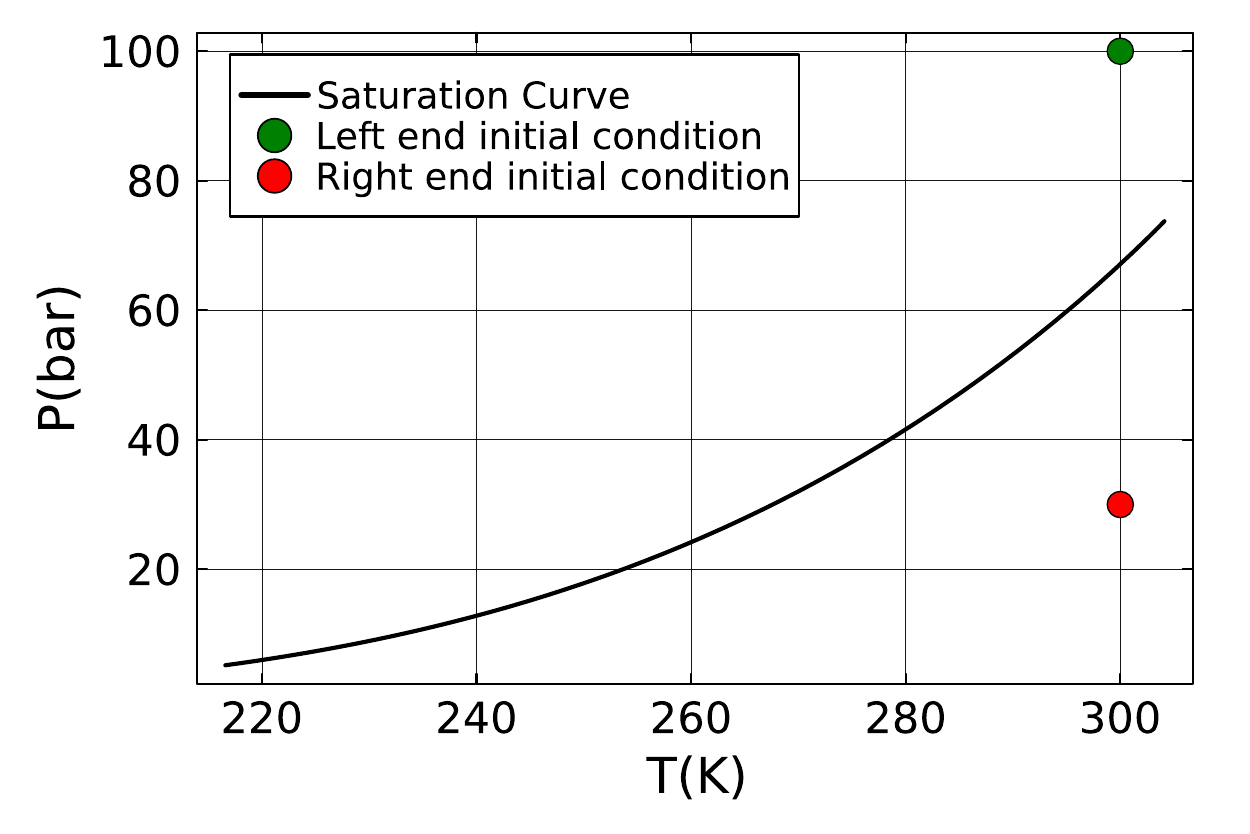}
    \caption{Initial conditions in $p-T$ space.}
    \label{fig:Initial_Conditions_PT}
\end{figure}

Table \ref{table:Pipleline-Sim-Params} details the initial conditions for the simulation, employing again \coo as the working fluid. Figure \ref{fig:Initial_Conditions_PT} shows the location of the left and right initial states with respect to the saturation curve. Initially, the fluid is at rest(i.e. u = 0). 
All simulations employ CFL values of 1.0 for algebraic approaches and 0.84 for the ODE approach. It is crucial to highlight that the ODE approach exhibits instability beyond a CFL value of 0.84 for this particular test case. Results are reported at \( t = 0.2 \) s. Given the discontinuous nature of initial conditions, an initial time-step of \( 1 \times 10^{-12} \) s is used to initiate the simulation.

The time step $\Delta t$ used in the simulations is based on the fastest traveling waves
\[\Delta t = \mathrm{CFL} \frac{\Delta x}{\underset{i}{\max} (\left| u_i \pm a_i \right|)}, \]
where $a_i$ represents the speed of sound, and subscript $i$ denotes the $i^{th}$ cell. 

\begin{table}[h!]
\centering
\caption{Pipeline simulation parameters based on \Hammer\ \cite{hammer_method_2014}. See also Figure \ref{fig:Initial_Conditions_PT}. \label{table:Pipleline-Sim-Params}}
\begin{tabular}{@{}ccccccc@{}}
\toprule
Simulation time (s) & Length (m) & Discontinuity location & \multicolumn{2}{c}{Left} & \multicolumn{2}{c}{Right} \\
           &               &                        & $p$ (bar) & $T$ (K) & $p$ (bar) & $T$ (K) \\ \midrule
0.2        & 200           & Centre                 & 100       & 300     & 30        & 300     \\ 
\bottomrule
\end{tabular}
\end{table}

\subsubsection{Validation with Hammer}
\begin{figure}[htbp]
     \begin{subfigure}[b]{0.48\textwidth}
         \includegraphics[width=\textwidth]{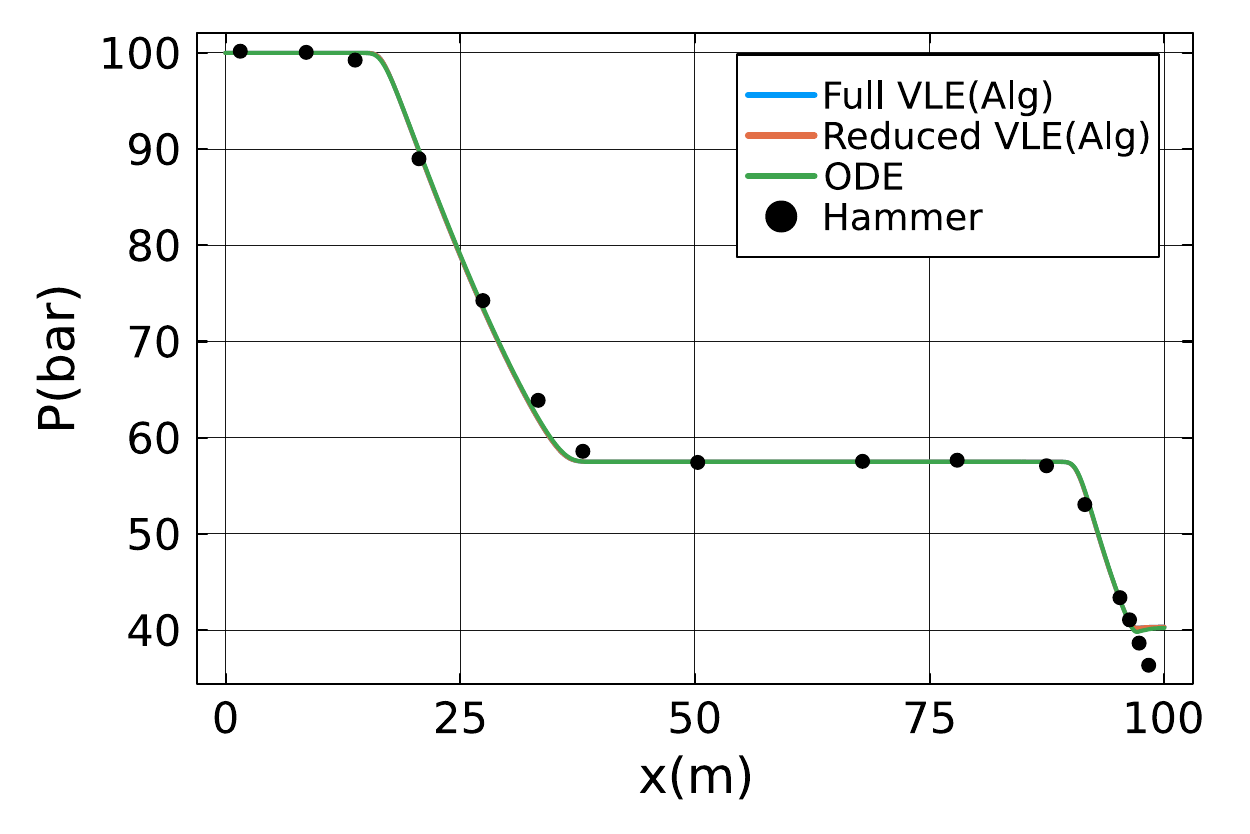}
         \caption{Comparison with \Hammer}
         \label{fig:ComparisonAllMethodsPipeWithHammer}         
    \end{subfigure}  
    \begin{subfigure}[b]{0.48\textwidth}
     \includegraphics[width=\textwidth]{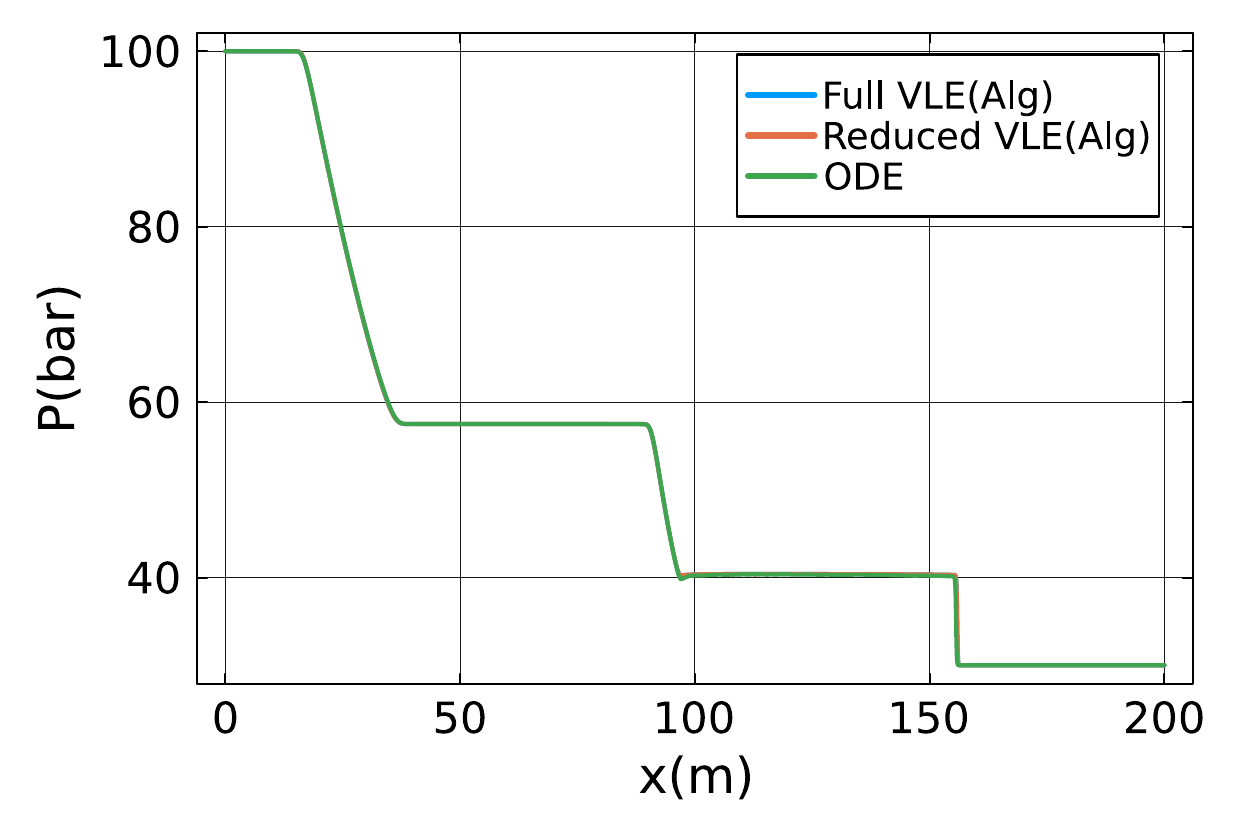}
      \caption{Comparison along full pipe}
      \label{fig:ComparisonAllMethodsPipeShockTube_1000_cells}           
\end{subfigure}
\caption{Pressure variation along pipe at $t = 0.2$\ s, with $N=4800$ cells. }
\end{figure}

First, we establish the validity and convergence of our solution. In Figure \ref{fig:ComparisonAllMethodsPipeWithHammer}, we validate our results with those from 
\Hammer\ \cite{hammer_method_2013}, which shows the results only for the left half of the pipe. A notable alignment is observed with \Hammer\ across all three approaches, albeit with a slight deviation at the right boundary. Additionally, all three approaches are nearly identical, overlapping significantly in the presented results. The discrepancy compared to \Hammer at the right boundary is expected as we use a different test case set-up, as explained before: a shock-tube experiment instead of an outflow boundary condition as used by \Hammer 
 
Figure \ref{fig:ComparisonAllMethodsPipeShockTube_1000_cells} presents a comparison of all three approaches for the entire length of the pipe. All approaches demonstrate good agreement, with nearly overlapping outcomes. It is important to note that the constraints are satisfied to machine precision in the algebraic methods. In contrast, the ODE approach involves computing the temperature for the subsequent time step via numerical integration, which introduces a truncation error of the order \( \mathcal{O}(\Delta t^2) \) compared to the algebraic method. This truncation error results in an energy drift error, as elaborated in the subsection \ref{pipe.subsec: energy-conservation-error}.

\begin{figure}[htbp]
     \centering
     \begin{subfigure}[b]{0.48\textwidth}
         \includegraphics[width=\textwidth]{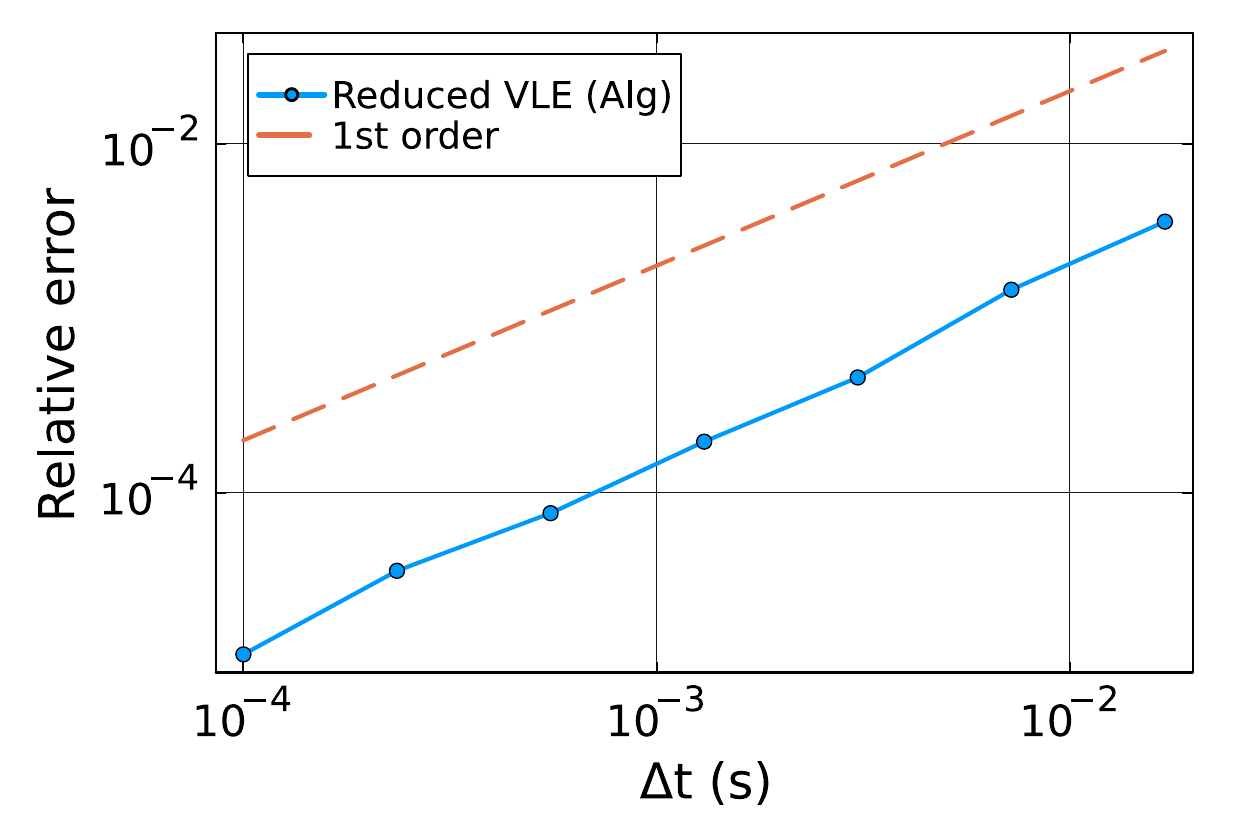}
         \caption{ \reducedDae}
         \label{fig:reducedVLETemporalConvergence}         
    \end{subfigure}  
    \begin{subfigure}[b]{0.48\textwidth}
     \includegraphics[width=\textwidth]{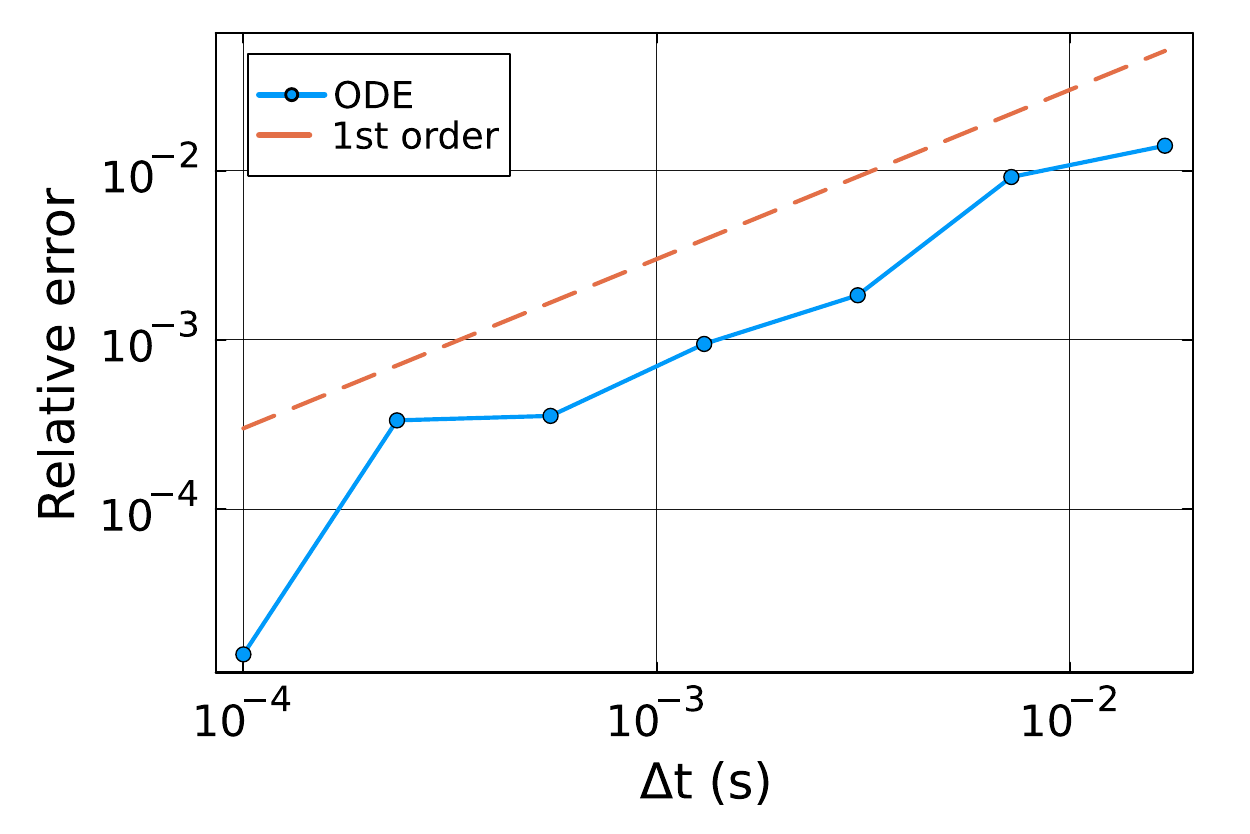}
      \caption{\ode.}
      \label{fig:reducedODETemporalConvergence}           
    \end{subfigure}
\caption{Temperature error: Results at $t=0.2$~s. 20 grid cells}
\end{figure}

\begin{figure}[htbp]
     \begin{subfigure}[b]{0.48\textwidth}
         \includegraphics[width=\textwidth]{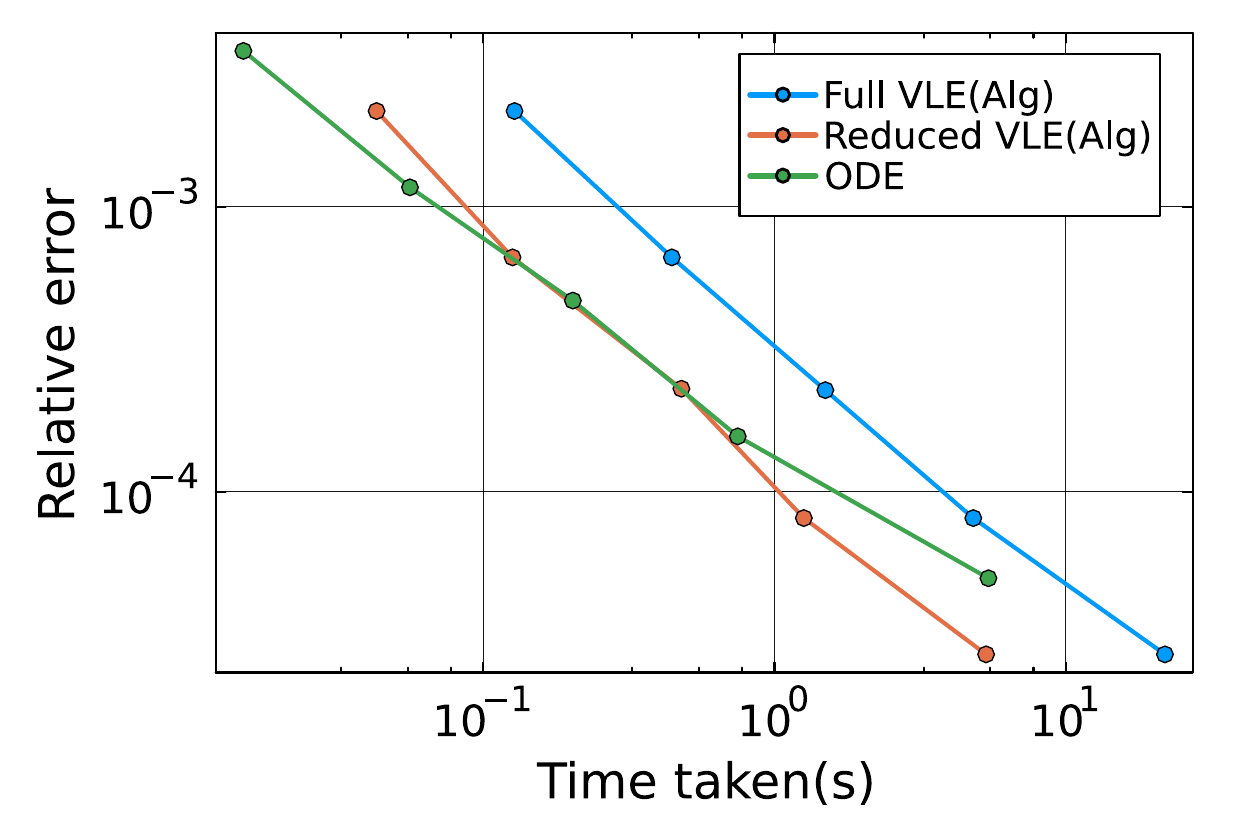}
         \caption{Pressure error}
         \label{fig:PressureErrorVsTimeTakenAllMethods}         
    \end{subfigure}  
    \begin{subfigure}[b]{0.48\textwidth}
     \includegraphics[width=\textwidth]{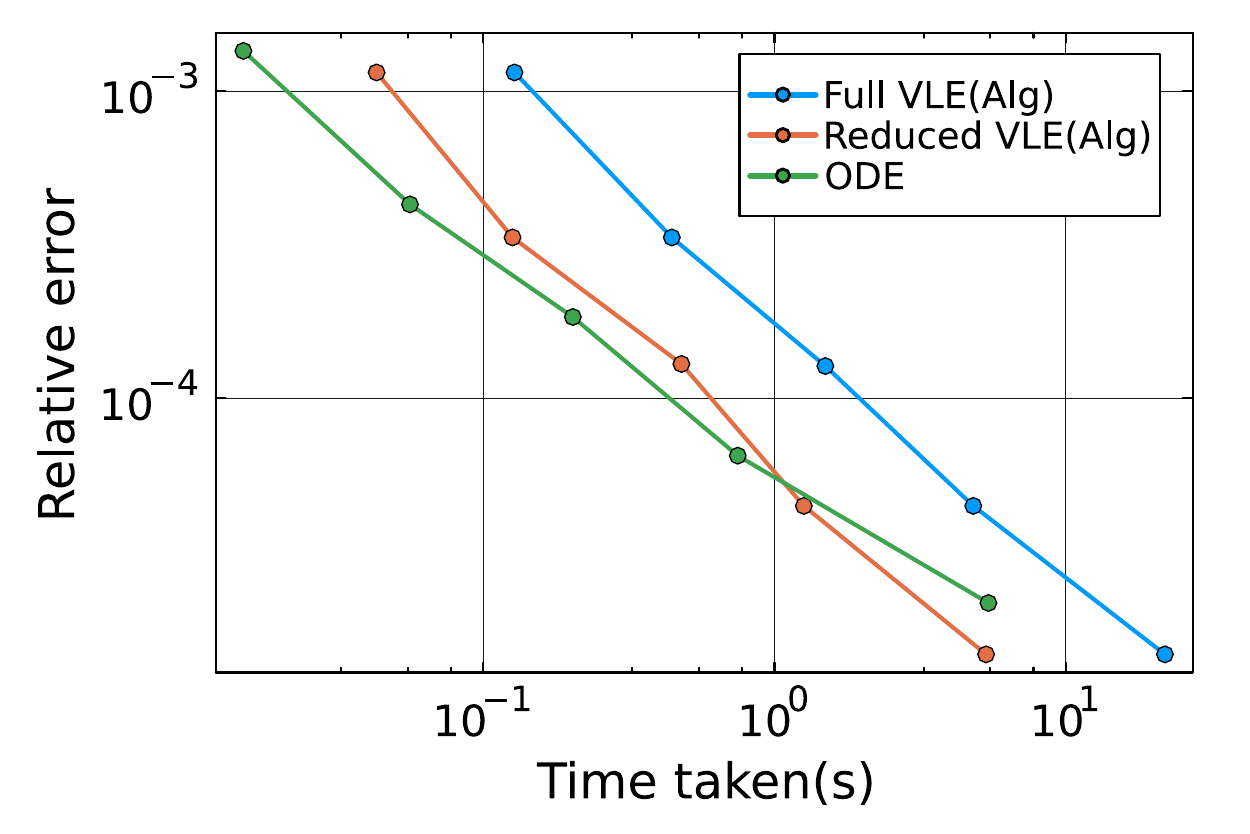}
      \caption{Temperature error}
      \label{fig:TemperatureErrorVsTimeTakenAllMethods}           
\end{subfigure}
\caption{Comparison of CPU time vs error in all approaches.}
\label{fig:ErrorVsTimeTakenAllMethods}
\end{figure}

\subsubsection{Convergence study}
We now investigate the spatial and temporal convergence of the new approaches. Spatial convergence is discussed in  \ref{app:spatial_convergence}. Here, we focus on temporal convergence. Figures \ref{fig:reducedVLETemporalConvergence} and \ref{fig:reducedODETemporalConvergence} depict the temporal convergence patterns observed in the \reducedDae and \ode methodologies, respectively, utilizing the $L_{1}$-norm for the error. The reference solution corresponds to the solution computed with a time step of $\Delta t = 5 \times 10^{-5}$~s. The error in temperature is calculated for each cell as $\varepsilon_T = (T - T_{\text{ref}})/T_{\text{ref}}$. Notably, the temporal convergence of the \reducedDae approach exhibits a highly favourable trend, demonstrating anticipated (first order) convergence characteristics whereas the \ode approach displays a near-first order convergence behavior. The observed outcome aligns with expectations, given that the \ode approach relies on the smoothness of the function $\f$ in the thermodynamic phase-plane ($\rho-T-\f$ space) which is a criterion that does not hold across phase transitions. This is particularly evident in the right-hand side of the temperature equation, which includes derivatives of $\f$. From a thermodynamic standpoint, it is known that across phase boundaries, the derivatives of internal energy exhibit discontinuities.

\subsubsection{Computational efficiency}
In this subsection, we discuss the computational efficiency of two new reduced methodologies. The \reducedDae and \ode approaches exhibit comparable performance. Both methods are significantly faster than the \fullDae approach, being approximately 3-4 times faster. This is evidenced in Figure~\ref{fig:ErrorVsTimeTakenAllMethods}, which presents plots of relative pressure and temperature error versus CPU time. The reference solution in this figure corresponds to the results obtained using 12000 cells with the \fullDae approach.

\subsubsection{Energy conservation error in \ode} \label{pipe.subsec: energy-conservation-error}
\begin{figure}[htbp]
    \centering
    \begin{subfigure}[b]{0.48\textwidth}
        \includegraphics[width=\textwidth]{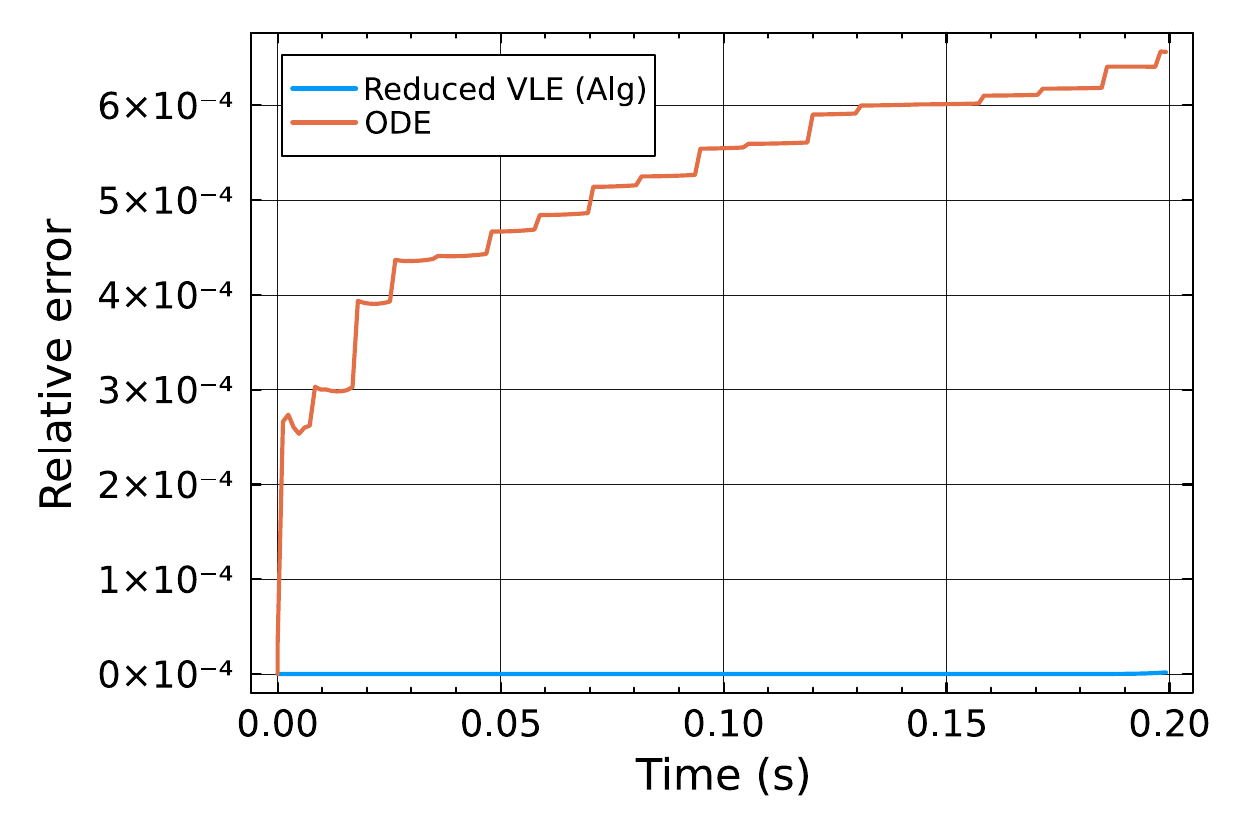}
        \caption{Error throughout the simulation in \reducedDae vs \ode. }
        \label{fig:EnergyErrorODEvsDAE}
    \end{subfigure}
    \hfill
    \begin{subfigure}[b]{0.48\textwidth}
        \includegraphics[width=\textwidth]{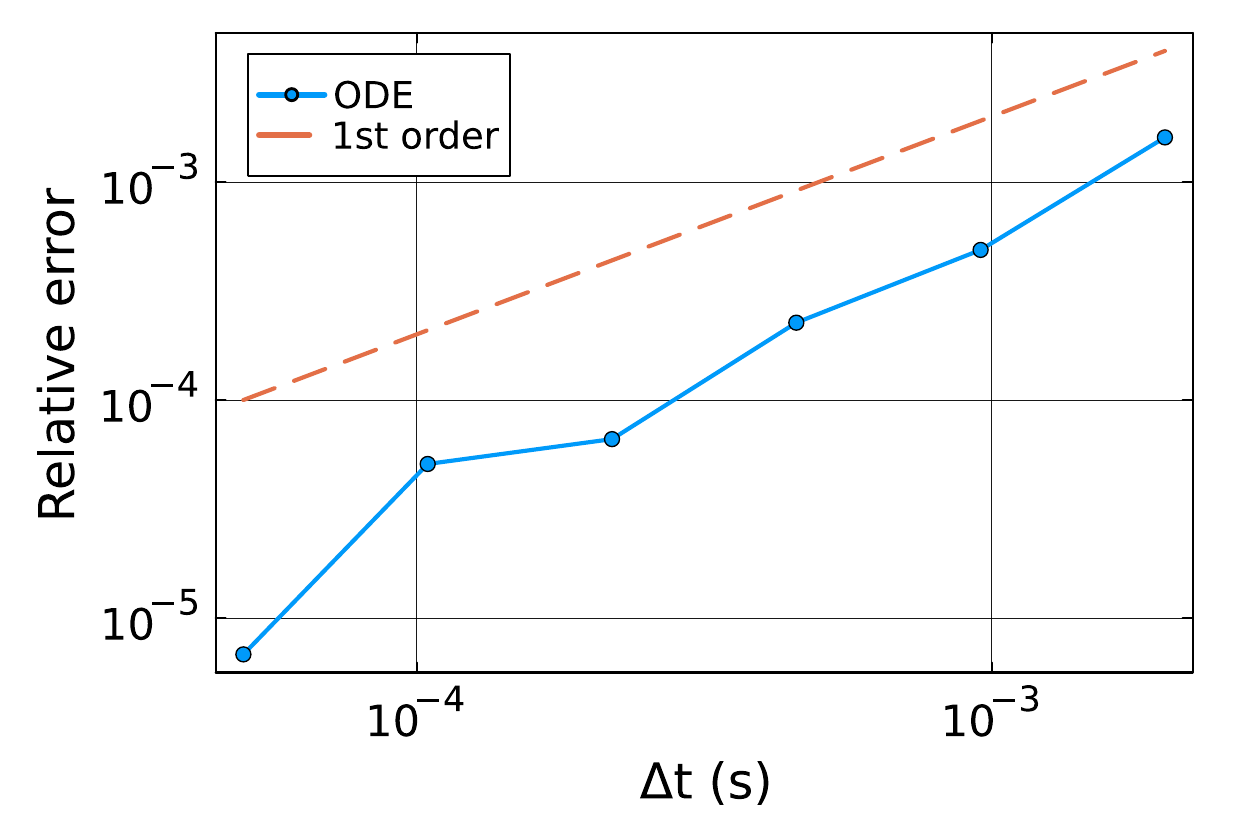}
        \caption{Error with varying time-steps in \ode.}
        \label{fig:reducedODETemporalConvergenceEnergy}
    \end{subfigure}
    \caption{Energy conservation error. Results at $t = 0.2$~s. 200 grid cells.}
    \label{fig:EnergyErrorTemporalConvergence}
\end{figure}

 Our investigation into the higher error observed in the \ode approach is further illustrated in Figure \ref{fig:EnergyErrorODEvsDAE} by presenting the relative energy error \( \varepsilon_{t, \rho E} \), which quantifies the total energy deviation throughout the entire pipe at time \( t \). The subscript \( 0 \) denotes the initial state of the pipe. The error is defined as:

\[
\varepsilon_{t, \rho E} = \frac{\sum_{i=1}^{N} \left(\rho_{t,i} E_{t,i} - \rho_{0,i} E_{0, i}\right)}{\sum_{i=1}^{N} \rho_{0,i} E_{0, i}},
\]
where \( N \) represents the total number of cells, and \( i \) denotes the cell index. This expression compares the total energy at time \( t \) to the initial energy state, offering a quantitative measure of the energy discrepancy between the current and initial states of the system. Both the \reducedDae and \ode approaches are evaluated based on this energy error metric. The \reducedDae approach inherently adheres to the energy conservation law, resulting in energy errors at machine precision. In contrast, the \ode approach bypasses direct use of the energy conservation equation. Instead, temperature is updated using the equation \eqref{pipe:temperature}, which introduces an error of order \odtt in temperature computation. This updated temperature is then used to calculate internal energy using the Span-Wagner EOS \cite{span_new_1996}, which consists of polynomial and exponential terms(with negative exponents). Errors in combined quantities \footnote{Combination implies performing an operation between two numbers/quantities e.g. addition, multiplication etc.} maintain their original order through basic arithmetic operations, implying that errors in internal energy are of the same order as those in density ($\rho$) and temperature ($T$).  Hence, in the total energy \eqref{pipe:total_energy_postprocessed}, we can expect an error of the order of \odtt. After a simulation time equal to $N \Delta t$ (where \(N\) is the number of timesteps), an accumulation of these errors can lead to an overall energy error of the order of \odt. This behaviour is confirmed by  Figure~\ref{fig:reducedODETemporalConvergenceEnergy}, which demonstrates an increase in energy drift with larger timesteps. It is noteworthy that the energy conservation error in both our \reducedDae and \ode approaches closely aligns with that reported in Sirianni et al. \cite{sirianni_explicit_2024}. The primary distinction between their numerical scheme and ours lies in the spatial discretization: while they utilize second-order spatial discretization, we employ only first-order spatial discretization. Additionally, it is pertinent to highlight that their work pertains to single-phase flow, whereas the emphasis of this paper lies in the realm of two-phase flow. 
Furthermore, our approach is based on the time-continuous form of the temperature equation, which offers more flexibility in employing different time integration methods. In contrast, their approach first discretizes in time using the forward Euler method.

We continue to describe the wave structure observed in the shock-tube results within the framework of the classical Riemann problem, with a specific emphasis on phase transition phenomena in the following subsection \cite{menikoff_riemann_1989}. 
\subsubsection{Wave structure}
\begin{figure}[htbp]
         \centering
         \includegraphics[width=\textwidth]{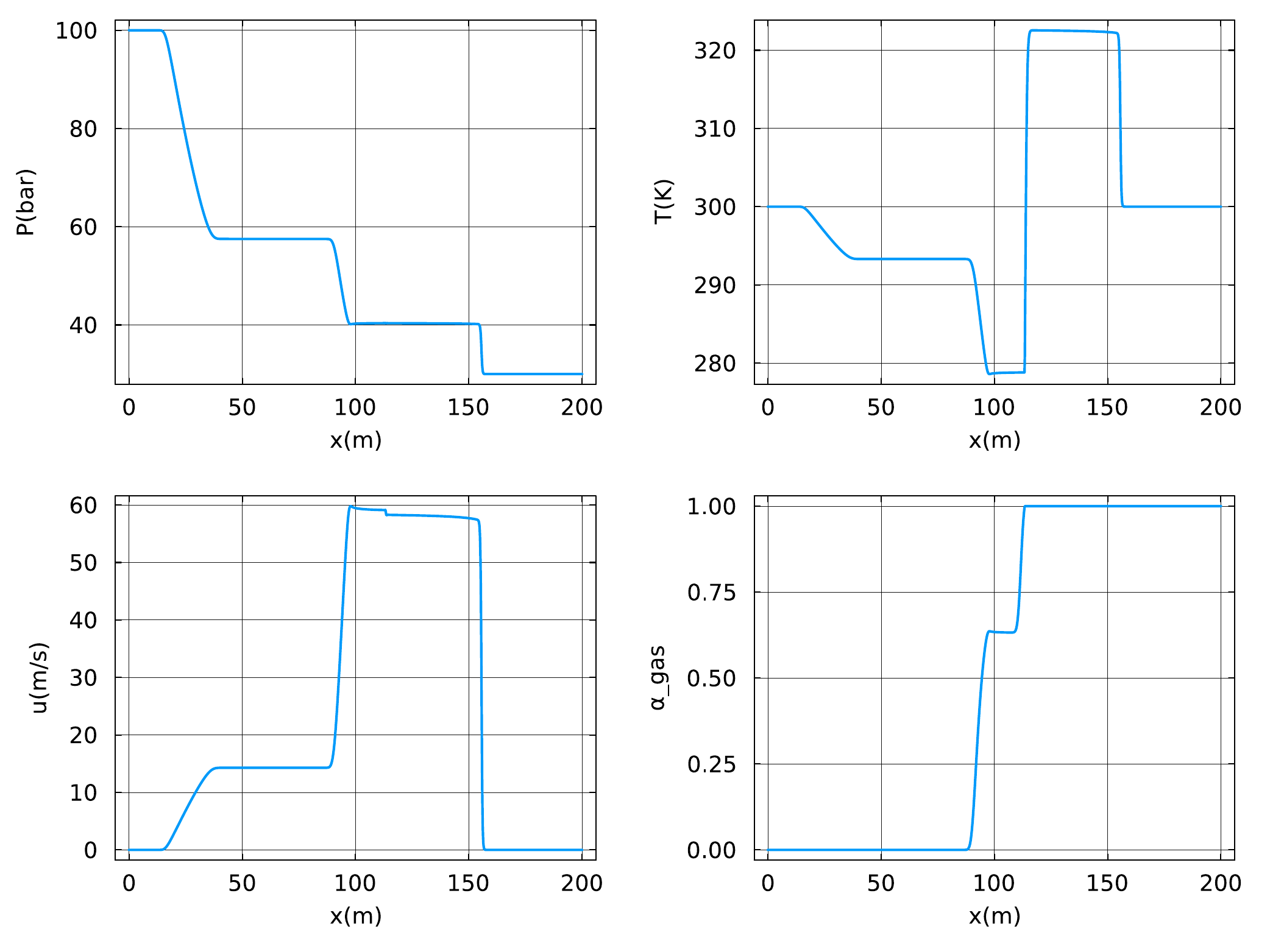}
         \caption{Riemann problem solution for various quantities along the length of the pipe for test case 1 with 1000 cells. Results at $t = 0.2$~s.}
         \label{fig:ShockTube_allVars_200m}
\end{figure}

In this subsection, we discuss the wave structure observed in our pipeline depressurization simulation. Figure~\ref{fig:ShockTube_allVars_200m} presents comprehensive results for test case 1 (Table~\ref{table:Pipleline-Sim-Params}) obtained using the \reducedDae method. These findings align well with observations by Hammer et al. (2013) \cite{hammer_method_2013}. The solution exhibits a rich wave structure, characterized by the presence of four distinct waves:

\begin{enumerate}
    \item Leftward-propagating rarefaction wave: This wave is characterized by a smooth decrease in both pressure and temperature as it propagates through the fluid.
    \item Evaporation wave: Marking the onset of boiling and the transition to two-phase \coo flow, the evaporation wave is accompanied by a further decrease in both pressure and temperature.
    \item Contact discontinuity: Separating the two-phase and single-phase gas regions of \coo, the pressure remains constant across the contact discontinuity, while temperature exhibits a significant rise.
    \item Rightward-moving shock wave: This wave induces compression that leads to increased internal energy, temperature, and pressure.
\end{enumerate}

\begin{figure}[htbp]
         \centering
         \includegraphics[width=\textwidth]{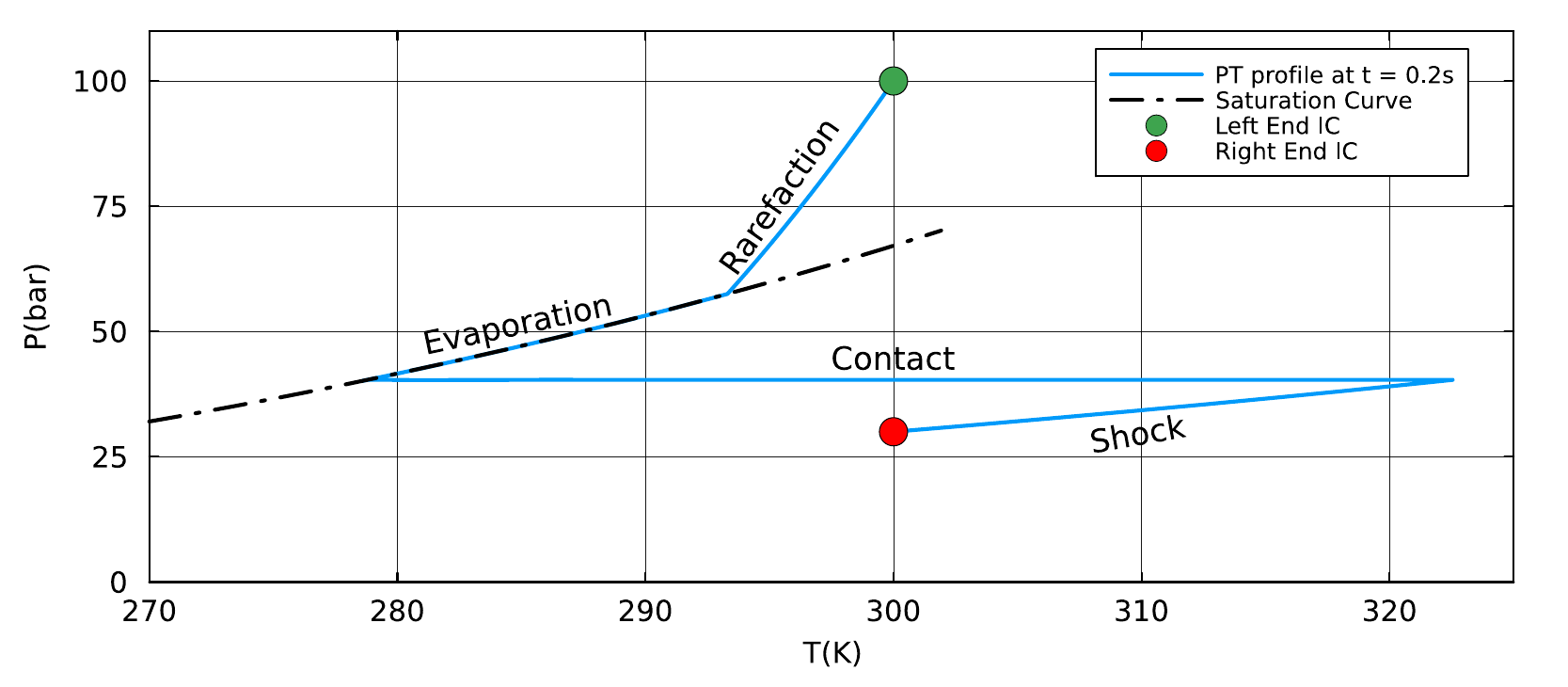}
         \caption{Pressure-temperature profile at $t=0.2s$ for test case 1 with 1000 cells.}
         \label{fig:PT_profile}
\end{figure}

Figure~\ref{fig:PT_profile} further visualizes these distinct waves in pressure-temperature space at the end of the simulation ($t = 0.2$\ s). The fluid initially transitions to a two-phase state via the rarefaction wave, followed by a further decrease in pressure-temperature properties through the evaporation wave. The contact discontinuity then separates the two-phase mixture from the hot, high-pressure single-phase gas. Finally, the shock wave reconnects the system to its initial state on the right end of the pipe.


\FloatBarrier
\section{Conclusions}\label{sec:conclusions}

In this paper, we have proposed two new numerical methods for simulating the depressurization of tanks and pipelines containing \coo. This involves solving fluid flow equations alongside thermodynamic equilibrium equations. 
A common approach is to advance the fluid equations while coupled with the non-linear algebraic constraints imposed by thermodynamic equilibrium. This approach, termed \fullDae, is computationally expensive. We propose two novel approaches by reformulating the coupled system of equations, which we denote \reducedDae and \ode. Firstly, the \reducedDae approach is based on the insight that the four-equation system expressing the thermodynamics constraint can be simplified to a one-equation constraint when using saturation relations in two-phase conditions. This accelerates the solution procedure at a slight loss of accuracy, which depends on the accuracy with which the saturation relations have been determined from the complete EOS. Secondly, the \ode approach is based on the insight that the one-equation constraint can be differentiated in time to yield an evolution equation for the temperature, which can be efficiently solved with explicit time integration methods, thus avoiding the need to employ a non-linear equation solver. However, this method introduces a small error in energy conservation. For simplicity, we have used the forward Euler method for time integration in all tests.

For the tank depressurization case, the two proposed methods are shown to be accurate and stable for different time-step sizes. The \reducedDae and \ode approaches exhibit comparable performance and we gain a significant speedup compared to \fullDae. The results are in excellent agreement with those reported in the literature. We also apply these methods to pipeline depressurization, where both \reducedDae and \ode again, exhibit significantly improved computational efficiency compared to the traditional approach. 

The pipeline depressurization case modelled as a Riemann problem, shows how four different waves appear: rarefaction, shock, contact discontinuity, and evaporation wave. All three methods effectively capture these intricate wave dynamics and results show excellent agreement with the literature.

One potential path to improve accuracy could be the incorporation of advanced time integration methods for systems with constraints into the current \ode framework.
Extending the methodology beyond single-component fluids to encompass multi-component mixtures presents another compelling direction for future work. This would necessitate addressing the complexities associated with multi-component systems, such as intercomponent mass transfer and the establishment of phase equilibria. Successfully navigating these challenges would broaden the applicability of the methodologies and enable the investigation of a wider range of fluid systems with enhanced accuracy and fidelity.

\section*{CRediT author statement}


\textbf{Pardeep Kumar}:
Conceptualization,
Methodology,
Software,
Validation,
Visualization,
Writing -- Original Draft

\textbf{Benjamin Sanderse}:
Project administration,
Formal analysis,
supervision,
writing - review \& editing

\textbf{Patricio I. Rosen Esquivel}:
Project definition,
funding acquisition,
supervision,
writing - review \& editing

\textbf{R.A.W.M. Henkes}:
Project administration,
Funding acquisition,
Writing -- Review \& Editing

\section*{Declaration of Generative AI and AI-assisted technologies in the writing process}


During the preparation of this work the authors used GitHub Copilot in order to
propose wordings and mathematical typesetting. After using this tool/service,
the authors reviewed and edited the content as needed and take full
responsibility for the content of the publication.

\section*{Declaration of competing interest}


The authors declare that they have no known competing financial interests or
personal relationships that could have appeared to influence the work reported
in this paper.

\section*{Acknowledgements}

This research was generously supported by Shell Projects and Technology, and we gratefully acknowledge their contribution. We also thank our colleagues Syver Døving Agdestein and Marius Kurz for their insightful discussions and constructive feedback.

\bibliographystyle{abbrv}
\bibliography{metadata/references}

\appendix
\section{Spatial convergence} \label{app:spatial_convergence}
\begin{figure}[htbp]
         \centering
         \includegraphics[width= 0.49\textwidth]{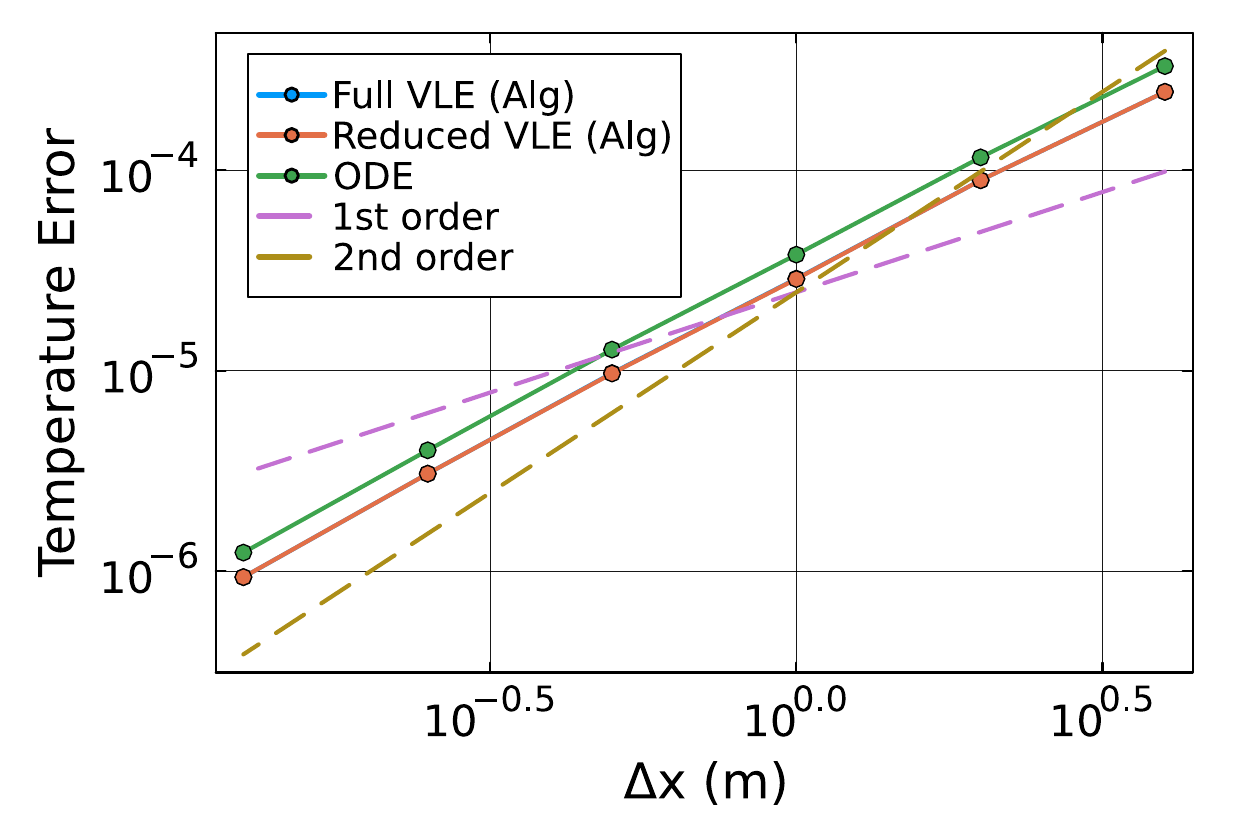}
         \caption{Pipeline depressurization: Spatial convergence along the pipe at $t = 0.2$\ s.}
         \label{fig:ConvergencePipeAllApproachesRev3}
\end{figure}

We now assess the spatial convergence for the pipeline simulation. 
The problem setup is discussed in section \ref{sec:pipe:problem_setup}. For this verification, we employ a high-fidelity reference solution obtained using a \fullDae approach with a very fine mesh of 12,000 cells. Figure \ref{fig:ConvergencePipeAllApproachesRev3} illustrates the spatial convergence of our results with mesh refinement, employing the $L_{1}$ error norm in temperature. All three methodologies showed increased accuracy as the mesh is refined. Convergence is achieved at a level between first and second order.  Notably, the plots corresponding to the \fullDae and \reducedDae approaches coincide, whereas the \ode approach exhibits slightly higher error as is investigated in the following section.
\section{Spatial discretization of HEM model} \label{HEM_HLLC}

To compute the numerical flux $\hat{\EulerFlux}_{i+\frac{1}{2}}$, the HLLC scheme, known as the Harten-Lax-van Leer-Contact scheme, is employed as an approximate Riemann solver \cite{toro_riemann_2009}. This scheme utilizes the wave structure inherent in the Riemann problem to estimate the flux at the interface between adjacent cells. In the context of single-phase flow, the Riemann solution typically features three distinct waves: rarefaction, shock, and contact waves. However, in the case of two-phase flow, an additional wave called the evaporation wave may be present. Leveraging the Rankine-Hugoniot condition, the HLLC scheme approximates the fluxes associated with each of these waves. The intermediate region between the rarefaction and shock waves is commonly referred to as the star-region. A comprehensive derivation of the HLLC scheme for Euler's equations is provided by Toro \cite{toro_riemann_2009}. Below, we present the expressions for the fluxes at the interface $i+\frac{1}{2}$:

\begin{align}
\hat{\EulerFlux}_{i+1/2} &= 
\begin{cases}
    \hllcF_L, & \text{if } 0 \leq S_L, \\
    \hllcF^*_{L}, & \text{if } S_L \leq 0 \leq S^*, \\
    \hllcF^*_{R}, & \text{if } S^* \leq 0 \leq S_R, \\
    \hllcF_R, & \text{if } 0 \geq S_R .
\end{cases} \\
\hllcF^*_{L} &= \hllcF_L + S_L (U^*_{L} - U_L), \\ 
\hllcF^*_{R} &= \hllcF_R + S_R (U^*_{R} - U_R), \\
S_L &= \min(u_L - a_L, u_R - a_R), \\
S_R &= \max(u_L + a_L, u_R + a_R), \\
S^\ast &= \frac{p_R - p_L + \rho_L u_L (S_L - u_L) - \rho_R u_R (S_R - u_R)}{\rho_L (S_L - u_L) - \rho_R (S_R - u_R)}, \\
U^*_{K} &= \rho_K \left( 
\frac{S_K - u_K}{S_K - S^*} \right) \begin{bmatrix}
    1 \\
    S^* \\
    E_K + (S^* - u_K) \left [ S^* + \frac{p_K}{\rho_K(S_K - u_K)}\right]
\end{bmatrix}.
\end{align}
Here, $K = L$(left state) or $K = R$(right state), $U_K \approx \cU(x_{K},t)$, $\hllcF_{K} = \EulerFlux(U_{K}), a_{K}$ is the speed of sound computed using the equation of state \cite{span_new_1996}. $S_L$ and $S_R$ are the wave speeds of the left and right-going waves, respectively, and $S^*$ denotes the speed of the contact wave. The accuracy of the HLLC method crucially depends on the precision of the estimates of the wave speeds $S_L$ and $S_R$. In this context, we have presented one method for estimating these wave speeds. For a comprehensive discussion of wave speed estimation, we refer the interested reader to Toro \cite{toro_riemann_2009}.

\subsection{Boundary conditions}
This subsection details the boundary conditions employed for the pipeline simulation test case, whose problem setup is described in section~\ref{sec:pipe:problem_setup}. Our initial condition will be such that we have choked flow conditions at the right end of the pipeline. So, boundary effects cannot propagate inside the computational domain. For the left end, we end our simulation before the fastest wave (the rarefaction wave) reaches there.

To study the evaporation wavefront in more detail (as described in \cite{saurel_modelling_2008}), the computational domain is extended. The simulation time is also adjusted to ensure that the fastest traveling waves do not reach the boundary of the computational domain.
In the right section of the pipe, we take atmospheric conditions as initial data.

\end{document}